\documentclass[aps,showpacs]{revtex4}
\usepackage{graphicx}
\usepackage{color}
\begin{document}

\title{Physics of ultracold Fermi gases revealed by spectroscopies}

\author{P\"aivi T\"orm\"a}

\affiliation{COMP Centre of Excellence, Department of Applied Physics, Aalto University, FI-00076 Aalto, Finland}

\begin{abstract}
This article provides a brief review of how various spectroscopies have been used to investitage many-body quantum phenomena in the context of ultracold Fermi gases. In particular, work done with RF spectroscopy, Bragg spectroscopy and lattice modulation spectroscopy is considered. The theoretical basis of these spectroscopies, namely linear response theory in the many-body quantum physics context is briefly presented. Experiments related to the BCS-BEC crossover, imbalanced Fermi gases, polarons, possible pseudogap and Fermi liquid behaviour and measuring the contact are discussed. Remaining open problems and goals in the field are sketched from the perspective how spectroscopies could contribute.
\end{abstract}

%
\maketitle

\vspace{2pc}
\noindent{\it email}: paivi.torma@aalto.fi

%
%
%

\section{Introduction}

One remarkable use of light, or radiation in general, is spectroscopy. Throughout the history of science, it has helped the humankind to discover, analyse, diagnose and understand. Here I discuss spectroscopies in one specific important context, namely ultracold Fermi gases. Ultracold Fermi gases have already enabled remarkable achievements, such as establishing experimentally the existence of the BCS-BEC crossover. In the year of light, 2015, it is timely to discuss how spectroscopies have contributed to these successes and how they could be used for solving remaining and new problems. It is timely also because the year 2015 is the tenth anniversary of fermionic superfluidity in ultracold gases. 

The aim here is to briefly review some of the most important topics explored by spectroscopies in the context of ultracold Fermi gases. I have already written a book chapter on the theory of spectroscopies in ultracold gases \cite{Torma2015}; there, the focus was on detailed explanation of the physics of the spectroscopies themselves, from the quantum many-body perspective, while the research done using them was hardly reviewed at all. Here the focus is exactly the opposite: the book chapter and this article are thus complementary and together form a substantial study material for anybody who wishes to start understanding the topic in depth. For a researcher who begins to use a certain method in lab or in calculations, it shoud be useful both to understand the theoretical basis as well as get inspiration from research problems the method has already been applied to. This gives the motivation for me to write both the book chapter and this article. In addition, although the focus on spectroscopies makes it somewhat incomplete and biased, this article serves also as a quick overview of the main physics explored with ultracold Fermi gases within the last then years. 
 
There are already several excellent review articles which discuss physics of ultracold Fermi gases; here a few comments about the overlap and the differences comparted to this article (and my related book chapter \cite{Torma2015}). Two famous reviews appeared in 2008, on many-body physics with ultracold gases by Bloch, Dalibard and Zwerger \cite{Bloch2008}, and on theory of ultracold atomic Fermi gases by  Giorgini, Pitaevskii and Stringari \cite{Giorgini2008}. Due to the vast scope of these reviews, the discussion on physics of ultracold Fermi gases revealed by spectroscopies is a bit more brief than presented here and in \cite{Torma2015}. This is the case also with some more recent reviews such as the one by Randeria and Taylor 2014 \cite{Randeria2014}. A classic book chapter on ultracold Fermi gases was written by Zwierlein and Ketterle in 2008 \cite{Ketterle2008}. Therein, presentation of spectroscopies and physics revealed by them is very thorough, however, only developments until mid 2008 are discussed. Ref.~\cite{Ketterle2008} has a substantial overlap with this article concerning the years 2003-2007, but for the sake of completeness, I wish to repeat the discussions of the early experiments. There is also a review focusing on the possibility of a pseudo-gap in ultracold Fermi gases by Chen, Levin and coworkers from 2009 \cite{Chen2009} which has overlap with related discussions here. Since 2009, reviews on specific topics within the research of ultracold Fermi gases have appeared: on density imbalanced gases by Sheehy and Radzihovsky \cite{Sheehy2007,Radzihovsky2010}, by Chevy and Mora \cite{Chevy2010}, and by Gubbels and Stoof \cite{Gubbels2013}, on lattice physics by Georges and Giamarchi \cite{Georges2012}, on polarons by Massignan, Zaccati and Bruun \cite{Massignan2014}, and on 2D Fermi gases by Levinsen and Parish \cite{Levinsen2015}. Some of these discuss particular spectroscopy experiments related to the topic of this article, and I will cite these in the sections below. This article presents briefly all major ultracold Fermi gas physics achievements where spectroscopies played a role during 2003-2015; it is therefore suited for a reader who wishes to get a quick glance through all the doors that spectroscopies have opened during the last decade.   

\begin{figure}
\includegraphics[width=0.7\textwidth]{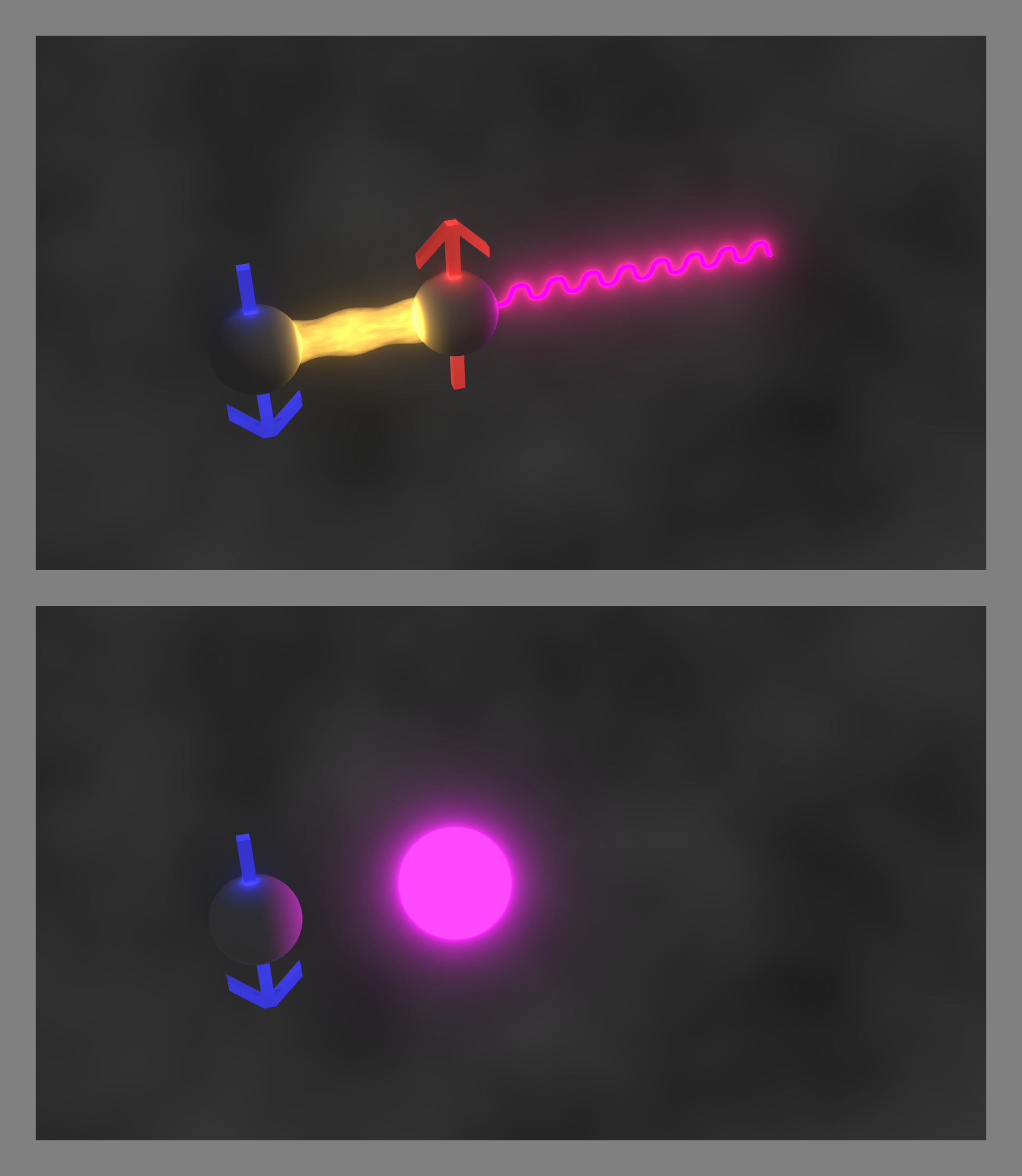}
\caption{Schematic of probing pairing by spectroscopies. The two fermions of different species (different pseudospins) correspond to either different internal states of an atom, or two different atomic species or isotopes. They form Cooper pairs or molecules due Feshbach-resonance enhanced interparticle interactions (yelllow). The probing field (the pink photon) transfers one of the species to a third one (another internal state, depicted by the pink sphere). The pair is broken. The energy needed for this process is shifted, due to the pairing energy, from the energy difference between the two internal states. This leads to spectral shifts that can be related to the properties of the molecule or the many-body state. Picture by Antti Paraoanu.} 
\label{SchematicTorma}
\end{figure}

I will first present the theoretical formalism of the three spectroscopies that are in focus: RF spectroscopy (see Figure \ref{SchematicTorma}), Bragg spectroscopy, and lattice modulation spectroscopy. I then proceed to describe the physics: how spectroscopies contributed to understanding issues such as BCS-type pairing and molecule formation, BCS-BEC crossover, the pairing gap, imbalanced Fermi gases and quasiparticle spectroscopy, the Mott-insulator phase for fermions in lattices, the contact, and the pseudogap vs. Fermi liquid nature of the normal state. Future expected uses of spectroscopies, for instance in probing the FFLO state, are outlined as well.

\section{The theoretical basis of the spectroscopies}

Spectroscopies are usually based on the principles of linear response. It is assumed that the probing field (light or other electromagnetic field) is a weak perturbation to the system, and the response thus represents basically properties of the probed system of interest, not the dynamics of the combined entity including the field and the system, c.f.\ Figure \ref{SchematicTorma}. Linear response theory often leads to the Fermi's golden rule, but in case of interacting many-body systems, one should be cautious to straighforwardly apply a simple Fermi golden rule formula. It is better to start from general linear response theory for many-body states, and carefully check which types of approximations are justified in the particular case of study. A detailed theoretical description suited for guiding such analysis is given in my book chapter \cite{Torma2015}. Here I will now present the theoretical basis of ultracold gas spectroscopies only briefly; for subtle details and further information please see Ref.\cite{Torma2015}. I will first discuss a typical system Hamiltonian for an ultracold Fermi gas, then present the first and second order linear response formulas in a general form, and finally discuss how they apply to RF, Bragg and lattice modulation spectroscopies.  

\subsection{The system and its Hamiltonian}

We are now interested in systems with several species of fermionic particles, possibly interacting with each other, and probed by a field.     
The fermions correspond to 
field operators $\hat{\psi}_\sigma(\bf{r})$ where $\sigma$ denotes the distinguishable species of particles, that is, (pseudo)spins (such as atoms in different internal states, or different atomic or molecular species). 
The Hamiltonian, in the second quantized field theoretical formalism, is
\begin{eqnarray}
\hat{H}_C&=&\int d^3r \sum_{\sigma}\hat{\psi}_{\sigma}^{\dagger}\left(\mathbf{r}\right)\left(-\frac{\hbar^{2}\nabla^{2}}
{2m_{\sigma}}
+E_{\sigma}+V_{T,\sigma}\left(\mathbf{r}\right)\right)\hat{\psi}_{\sigma}\left(\mathbf{r}\right) \nonumber \\ 
&+&\frac{1}{2}\sum_{\alpha,\beta}\int d^3r \int d^3r' V_{\alpha\beta}\left(\mathbf{r},\mathbf{r}'\right)\hat{\psi}_{\alpha}^{\dagger}\left(\mathbf{r}\right)\hat{\psi}_{\beta}^{\dagger}\left(\mathbf{r'}\right)\hat{\psi}_{\beta}\left(\mathbf{r'}\right)\hat{\psi}_{\alpha}\left(\mathbf{r}\right) \nonumber \\
&+& \sum_{\gamma,\delta}\int d^3r \left[ \hbar \Omega_{\gamma\delta} (\mathbf{r},t) \hat{\psi}_{\gamma}^{\dagger}\left(\mathbf{r}\right)\hat{\psi}_{\delta}\left(\mathbf{r}\right) + h.c. \right]. 
 \label{EqHPtorma}\end{eqnarray}
Here $m_{\sigma}$, $E_{\sigma}$ and $V_{T,\sigma}\left(\mathbf{r}\right)$ are the mass, internal state energy and the 
external (trapping) potential for
the different species, $V_{\alpha\beta}\left(\mathbf{r},\mathbf{r}'\right)$ is the interparticle interaction potential, 
and $\Omega_{\gamma\delta} (\mathbf{r},t)$ is the field that couples two species (the species must now be two internal states). Here $\Omega_{\gamma\delta} (\mathbf{r},t) = 
-\mathbf{d}_{\gamma\delta}\cdot \mathcal{E}(\mathbf{r},t)/\hbar$, where $\mathbf{d}_{\gamma\delta}$ is the transition dipole moment between the two internal states and $\mathcal{E}(\mathbf{r},t)$ is the electric field amplitude.
The sets of species that the indices $\alpha,\beta$ and $\gamma,\delta$ correspond to need not be (and usually are not) 
the same: there can be non-interacting species coupled by the field, or interacting species that do not interact with any 
fields. This Hamiltonian describes the full many-body quantum 
physics of the particles and the interaction with the field, albeit the field is assumed to be classical. Note that the 
chemical potentials $\mu_{\sigma}$ do not appear here. This means that the Hamiltonian is canonical (reason for the notation 
$\hat{H}_C$). This is because the last line of the Hamiltonian (\ref{EqHPtorma}) does not conserve particle number and thus 
chemical potential is not a meaningful concept. 

For calculating the response of the system to the external field, it is convenient to divide the Hamiltonian (\ref{EqHPtorma}) so that 
$\hat{H}'_L\left(t\right)$ denotes the last line and $\hat{H}_{0C}$ the rest.      

\subsection{Linear response for many-body quantum states} \label{basicLinRespTorma}

We now proceed to give the linear response formulas that describe the spectroscopic response. For that purpose, let us make the Hamiltonian (\ref{EqHPtorma}) slightly more specific.

The field is coupled to a transition between two internal states, which in ultracold gases may mean between two species (pseudospins) of the system, denoted
$e$ and $g$. Here the state $g$ does not need to be the ground state but is chosen to be lower in energy than the state $e$. 
In the Hamiltonian (\ref{EqHPtorma}) other species and indices than $e$ and $g$ may also contribute since it 
can be a Hamiltonian for a complicated many-body system. We now say that the field couples only two states which leads the last line of the Hamiltonian
having $\gamma = e$ and $\sigma = g$ and no summation:
\begin{eqnarray}
\hat{H}_C&=&\int d^3r \sum_{\sigma}\hat{\psi}_{\sigma}^{\dagger}\left(\mathbf{r}\right)\left(-\frac{\hbar^{2}\nabla^{2}}
{2m_{\sigma}}
+E_{\sigma}+V_{T,\sigma}\left(\mathbf{r}\right)\right)\hat{\psi}_{\sigma}\left(\mathbf{r}\right) \nonumber \\ 
&+&\frac{1}{2}\sum_{\alpha,\beta}\int d^3r \int d^3r' V_{\alpha\beta}\left(\mathbf{r},\mathbf{r}'\right)\hat{\psi}_{\alpha}^{\dagger}\left(\mathbf{r}\right)\hat{\psi}_{\beta}^{\dagger}\left(\mathbf{r'}\right)\hat{\psi}_{\beta}\left(\mathbf{r'}\right)\hat{\psi}_{\alpha}\left(\mathbf{r}\right) \nonumber \\
&+& \int d^3r \left[ \hbar \Omega (\mathbf{r},t) \hat{\psi}_{e}^{\dagger}\left(\mathbf{r}\right)\hat{\psi}_{g}\left(\mathbf{r}\right) + h.c. \right]. 
 \label{EqH2Ptorma}
\end{eqnarray}
The response is obtained by calculating the rate of change in one of the species (for instance $e$), 
$\hat{\dot n}_e = \frac{d}{dt} \left(\hat{\psi}_e^{\dagger}\left(\mathbf{r}\right)
\hat{\psi}_e\left(\mathbf{r}\right)\right)$. 
Basically the same results could be obtained from $\hat{n}_e$ instead of 
$\hat{\dot n}_e$; the difference is similar to calculating the transition probablity or the transition rate. 
Here $\hat{\dot n}_e$
is considered since it is the same as current in electrical transport (c.f.\ the so-called Kubo formulas, described, for instance, in Ref.~\cite{Mahan00} Section 3.8). In the context of
Bragg spectroscopy, however, we will consider $\hat{n}$. 

We give now the general linear response result for any
observable (operator) $\hat{O}$. Perturbation theory is done with respect to the perturbation part of the
Hamiltonian $\hat{H}'_L$ (same as the last line of (\ref{EqH2Ptorma})) and the rest of the Hamiltonian is $\hat{H}_{0C}$. 
In the interaction picture one has: $\hat{\psi}_{\sigma C}^{\dagger}\left(\mathbf{r}
,t\right) = e^{i\hat{H}_{0C} t/\hbar} \hat{\psi}_{\sigma}^{\dagger}\left(\mathbf{r}\right) e^{-i\hat{H}_{0C} t/\hbar}$ 
and 
$\hat{H}_{LC}\left(\mathbf{r},t\right) = e^{i\hat{H}_{0C} t/\hbar} \hat{H}'_L\left(\mathbf{r},t\right) 
e^{-i\hat{H}_{0C} t/\hbar}
\equiv \hat{H}_L(t)$.  
The time-evolution of the
state $| \Psi \rangle$ becomes $| \Psi (t)\rangle = U(t,t_0)| \Psi (t_0)\rangle $,
where $U(t,t_0)=T \exp (-\frac{i}{\hbar}\int_{t_0}^t dt' \hat{H}_L(t'))= 
1 -\frac{i}{\hbar}\int_{t_0}^t dt' \hat{H}_L(t') - \frac{1}{\hbar^2} \int_{t_0}^t dt' \int_{t_0}^{t'} dt'' \hat{H}_L(t')\hat{H}_L(t'') + {\cal O}(\hat{H}_L^3)$. 
Now, one can approximate $\langle \Psi | \hat{O}\left(\mathbf{r},t\right)  | \Psi \rangle$
by including only the first two terms of the expansion, giving the constant 
$\langle \Psi (t_0) |\hat{O} | \Psi (t_0)\rangle$ and terms
linear (first order) in $\hat{H}_L$. The first order term finally is
(the notation $|\rangle = | \Psi (t_0)\rangle$ is used)
\begin{eqnarray}
&&\langle \Psi (t) | \hat{O}\left(\mathbf{r},t\right)  | \Psi (t) \rangle_{1st}
=  - \frac{i}{\hbar} \langle \Psi (t_0) |\int_{t_0}^t dt' \left[ \hat{O}\left(\mathbf{r},t\right),
  \hat{H}_L(t') \right] | \Psi (t_0)\rangle \nonumber \\
&& = - \frac{i}{\hbar} \int_{t_0}^t dt' 
\langle \left[ \hat{O}\left(\mathbf{r},t\right),
  \hat{H}_L(t') \right]\rangle \nonumber \\
&& \equiv   \int_{t_0}^\infty dt' \chi(t,t') , \label{GeneralFullCurrentTorma}
\end{eqnarray}
where the first order susceptibility was introduced:
\begin{eqnarray}
\chi(t,t') = \frac{1}{i\hbar} \theta (t-t')
\langle \left[ \hat{O}\left(\mathbf{r},t\right),
  \hat{H}_L(t') \right]\rangle .
\end{eqnarray}
Putting together terms of second order in $\hat{H}_L$ leads to
\begin{eqnarray}
&&\langle \Psi (t) | \hat{O}\left(\mathbf{r},t\right)  | \Psi (t) \rangle_{2nd}
= - \frac{1}{\hbar^2} \left( \int_{t_0}^t \int_{t_0}^{t'} dt'dt''  
\langle \hat{H}_L(t'')\hat{H}_L(t')\hat{O}\left(\mathbf{r},t\right) \rangle \right. \nonumber \\
&& -  \int_{t_0}^t \int_{t_0}^{t} dt'dt'' \langle \hat{H}_L(t') 
\hat{O}\left(\mathbf{r},t\right)\hat{H}_L(t'') \rangle  \nonumber \\
&& \left. + \int_{t_0}^t \int_{t_0}^{t'} dt'dt''\langle \hat{O}\left(\mathbf{r},t\right)
 \hat{H}_L(t')\hat{H}_L(t'') \rangle \right) \nonumber \\
&& \equiv  \int_{t_0}^\infty \int_{t_0}^\infty dt' dt'' \chi(t,t',t'') , \label{General2ndFullCurrentTorma}
\end{eqnarray}
now with the second order susceptibility:
\begin{eqnarray}
\chi(t,t',t'') = \frac{\theta (t-t')\theta (t'-t'')}{(i\hbar)^2} 
\langle \left[\left[  \hat{O}\left(\mathbf{r},t\right),
  \hat{H}_L(t') \right], \hat{H}_L(t'')\right] \rangle .
\end{eqnarray}

\subsection{Including the chemical potential and making the rotating wave approximation} \label{RWATorma}

The field amplitude $\Omega\left(\mathbf{r},t\right)$ contains the constant Rabi frequency $\Omega$ and the time and spatial dependence 
of the field. In case of electromagnetic fields, the time dependence
is harmonic, for instance $\sin(\omega_L t)$. We proceed to do the standard rotating wave approximation where only the 
resonant terms $e^{i(\omega_L-\omega_{eg})t}$ are kept and the non-resonant $e^{i(\omega_L+\omega_{eg})t}$ 
are removed \cite{Grynberg10}. For that purpose, let us consider 
$\hat{\psi}_{\sigma}^{\dagger}\left(\mathbf{r}
,t\right) = e^{i\hat{H}_{0C} t/\hbar} \hat{\psi}_{\sigma}^{\dagger}\left(\mathbf{r}\right) e^{-i\hat{H}_{0C} t/\hbar}$.
One can add to the Hamiltonian $\hat{H}_{0C}$ the term $\sum_\sigma \mu_\sigma  
\hat{\psi}_{\sigma}^{\dagger}\left(\mathbf{r}\right) \hat{\psi}_{\sigma}\left(\mathbf{r}\right)$ and also subtract
it. The added chemical potential, as well as terms proportional to the internal
state energy explicitly are then written in the following way
\begin{equation}
\hat{\psi}_{\sigma}^{\dagger}\left(\mathbf{r}
,t\right) = e^{i\hat{H}_{0C} t/\hbar} \hat{\psi}_{\sigma}^{\dagger}\left(\mathbf{r}\right) e^{-i\hat{H}_{0C} t/\hbar}
= e^{i (E_\sigma + \mu_\sigma)t/\hbar} e^{i\hat{H}_{0} t/\hbar} \hat{\psi}_{\sigma}^{\dagger}
\left(\mathbf{r}\right) e^{-i\hat{H}_{0} t/\hbar} ,
\end{equation} 
where 
\begin{eqnarray}
\hat{H}_{0} &=&\int d\mathbf{r}\sum_{\sigma}\hat{\psi}_{\sigma}^{\dagger}\left(\mathbf{r}\right)
\left(-\frac{\hbar^{2}\nabla^{2}}{2m_{\sigma}}
-\mu_\sigma+V_{T,\sigma}\left(\mathbf{r}\right)\right)\hat{\psi}_{\sigma}\left(\mathbf{r}\right) \nonumber \\ 
&+&\frac{1}{2}\sum_{\alpha,\beta}\int d\mathbf{r}\int d\mathbf{r'}V_{\alpha\beta}\left(\mathbf{r},\mathbf{r}'\right)
\hat{\psi}_{\alpha}^{\dagger}\left(\mathbf{r}\right)\hat{\psi}_{\beta}^{\dagger}\left(\mathbf{r'}\right)
\hat{\psi}_{\beta}\left(\mathbf{r'}\right)\hat{\psi}_{\alpha}\left(\mathbf{r}\right). \label{MBHamTorma}
\end{eqnarray}
Here $\hat{H}_0$ is the many-body Hamiltonian without the field, but containing the chemical
potentials. This transformation is applied to modify the observable $\langle \hat{\dot{N}}_e\rangle$: terms of the form 
$e^{\pm i (E_e - E_g + \mu_e - \mu_g)t/\hbar}$ are obtained. These multiply the
$\sin (\omega_L t)= (e^{i\omega_L t} - e^{-i \omega_L t})/(2i)$ terms and now it is possible to take the usual rotating 
wave approximation, that is, keep only 
terms including $\omega_L - (E_e - E_g)/\hbar = \delta$ and ignore those with $\omega_L + (E_e - E_g)/\hbar$. The
chemical potential difference $\mu_g - \mu_e$ now has the same role as the detuning $\delta$. We therefore define
the generalized detuning $\tilde{\delta} = \delta + (\mu_g - \mu_e)/\hbar$. The particle number current becomes now 
\begin{eqnarray}
\langle \Psi (t) | \hat{\dot N}_e\left(t\right)  | \Psi (t) \rangle
&=& - \frac{1}{4} \int_{t_0}^t dt' \int d^3r \int d^3r' \Omega\left(\mathbf{r}\right) \Omega^*\left(\mathbf{r'}\right)  \label{MBcurrentTorma} \\
&&e^{- i \tilde{\delta}(t-t')}
\langle \left[ 
\hat{\psi}_e^{\dagger}\left(\mathbf{r},t\right)\hat{\psi}_{g}\left(\mathbf{r},t\right),
\hat{\psi}_g^{\dagger}\left(\mathbf{r'},t'\right)\hat{\psi}_{e}\left(\mathbf{r'},t'\right) \right] \rangle  + h.c. ,
\nonumber
\end{eqnarray}
where $\hat{\psi}_{g/e}\left(\mathbf{r},t\right)$ are operators transformed by the many-body Hamiltonian $\hat{H}_{0}$.

The spectroscopies considered in this article differ basically in two ways: first, which are the species
$e$ and $g$ that are coupled by the fields, and second, what is the $\mathbf{r}$-dependence of the coupling $\Omega \left(\mathbf{r}\right)$. The spatial dependence determines whether the field gives momentum to the system. These issues will be specified in the sections below.

\subsection{Linear response with mean-field approximation} \label{simpleLinearResponseTorma}

The equation (\ref{MBcurrentTorma}) is a general linear response result, with no assumptions about  
the many-body state but neglecting correlations containing terms of the type $\hat{\psi}_\sigma \hat{\psi}_\sigma$ \cite{Torma2015}. 
Further closed formulas for the response can be calculated when
one of the states, say $e$, does not interact with the state $g$ by collisions or similar. That is,
$e$ and $g$ are only coupled by the field. Then the state of the system before the perturbation
is a product of states for the species $e$ and $g$ and the four-operator correlators can be factorized to parts 
that contain only $e$ or $g$ operators. In other words,
\begin{eqnarray}
&&\langle \left[ 
\hat{\psi}_e^{\dagger}\left(\mathbf{r},t\right)\hat{\psi}_{g}\left(\mathbf{r},t\right),
\hat{\psi}_g^{\dagger}\left(\mathbf{r'},t'\right)\hat{\psi}_{e}\left(\mathbf{r'},t'\right) \right] \rangle  =  \label{FactorizedCurrentTorma} \\
&&\langle  
\hat{\psi}_e^{\dagger}\left(\mathbf{r},t\right)\hat{\psi}_{e}\left(\mathbf{r'},t'\right) \rangle 
\langle \hat{\psi}_{g}\left(\mathbf{r},t\right) 
\hat{\psi}_g^{\dagger}\left(\mathbf{r'},t'\right)\rangle 
 - 
\langle \hat{\psi}_e\left(\mathbf{r'},t'\right)\hat{\psi}^\dagger_{e}\left(\mathbf{r},t\right) \rangle 
\langle \hat{\psi}^\dagger_{g}\left(\mathbf{r'},t'\right) 
\hat{\psi}_g\left(\mathbf{r},t\right)\rangle . \nonumber
\end{eqnarray}
There are, however, cases when this a factorization is not justifiable, for instance if the final state $e$ interacts strongly with some of the initial states. In that case  evaluating the whole four-operator correlator, for instance using self-consistent schemes, is the way to approach.
 
\subsection{Response for initial BCS state and final normal state}

As an illustrative example, let us consider that the system is intially in the BCS state described by the Bogoliubov transformation \cite{deGennes99} which has new operators
$\hat{\gamma}$ that correspond to the quasiparticles in the BCS state. The relation to the original operators (with the choice that the species $g$ is initially in the BCS state): $\hat{c}_{lg} = 
u_l  \hat{\gamma}_{lg} + v_l \hat{\gamma}^\dagger_{lg'}$, where $u_l$ and $v_l$ are the Bogoliubov coefficients and $g'$ is the species
with which $g$ is paired. Here $l$ is a generic index referring to momentum or other quantum number, such as the trap quantum number. Assume also that final state in the spectroscopy is non-interacting. Now we apply this simple mean-field theory to the linear response formula, doing the approximation to the factorized form as in Equation (\ref{FactorizedCurrentTorma}). 
The current, for
species $e$ is in the normal state and $g$ in the BCS one, becomes
\begin{eqnarray}
I(\tilde{\delta}) 
&=& \frac{\pi}{2} \sum_{k,l} \left| \int d^3r' \Omega\left(\mathbf{r}\right)  
\varphi^*_{ke}({\bf r})\varphi_{lg}({\bf r'}) \right|^2 \nonumber \\
&&\times(-u_l^2 n_F(E_{ke})n_F(E_{lg}) \delta((E_{ke}+E_{lg})/\hbar-\tilde{\delta}) \nonumber \\
&&-v_l^2 n_F(E_{ke})(1 - n_F(E_{lg})) \delta((E_{ke}-E_{lg})/\hbar-\tilde{\delta}) \nonumber \\
&&+u_l^2 (1-n_F(E_{ke}))(1-n_F(E_{lg})) \delta((E_{ke}+E_{lg})/\hbar-\tilde{\delta}) \nonumber \\
&&+v_l^2 (1-n_F(E_{ke}))n_F(E_{lg}) \delta((E_{ke}-E_{lg})/\hbar-\tilde{\delta}))  .   \label{ManyBCurrTorma}
\end{eqnarray}          
The discussion so far has assumed zero temperature. 
Finite temperatures are often treated with the so-called Matsubara formalism, based on Green's functions. Then the zero-temperature 
Fermi distributions $n_F(E)$ above are simply the finite temperature distributions. In the following, I give
finite temperature results, unless noted otherwise. Here $\varphi_{k\sigma}({\bf r})$ is the basis wave function related to the quantum number $k$ (for instance plane waves or harmonic trap eigenfunctions) and $E_{k\sigma}$ are the quasiparticle energies labelled by $k$ and $\sigma$.

\subsection{RF spectroscopy} \label{RFTorma}

In RF spectroscopy, one applies an RF field to an atom or molecule that has a 
suitable transition between two of its internal states, say $g$ and $e$. The total system may have more species:
the $e$, the $g$ and some others (atoms in some other internal states). Often one of the species, say $e$, is not
present before applying the field: this is the final state and $g$ is the initial state. 
Note that the wavelength of the RF field is large compared to other relevant length scales, and is usually much larger than the 
cloud of atoms. Due to this fact, the momentum of the RF photon is negligible compared to, for example, typical scales of the
Fermi momentum, thus the photon momentum $k_L$ can be set to zero. Furthermore, the intensity of the RF field over 
the cloud can be assumed to be constant in the scale of the cloud. This causes the coupling parameter to be simply a constant 
$\Omega(\mathbf{r}) = \Omega \exp^{i \mathbf{k}_L \cdot \mathbf{r}}= \Omega$. Consequently, momentum conservation is imposed between 
the initial and final momenta of the particle that is transferred from the state $g$ to $e$, as seen from the overlap integrals
in Equation (\ref{ManyBCurrTorma}). The momentum of the final particles may or may not be resolved, depending on the case. 
   
The version of RF spectroscopy where momentum is not resolved is, for historical reasons, the "usual" or "standard" RF spectroscopy. If both $e$ and $g$ are in a normal 
state the current is
\begin{eqnarray}
I(\tilde{\delta})
&=& \frac{\pi}{2} \sum_{k} \left| \Omega \right|^2 
[n_F(E_{kg})(1 - n_F(E_{ke})) 
- n_F(E_{ke})(1 - n_F(E_{kg}))] \nonumber \\
&& \times \delta ((E_{ke}-E_{kg})/\hbar-\tilde{\delta})  . \label{RFnormalTorma}
\end{eqnarray} 
The momentum summation can be changed into energy integration, which brings the density of states to this 
formula and then the connection to Fermi's golden rule is clear. For instance 
energy shifts caused by interactions can be observed with RF spectroscopy. This is because, if the kinetic and potential energy terms in $(E_{ke}-E_{kg})/\hbar-\tilde{\delta}$ cancel, one is left with
$(E_{\mathrm{int},ke}-\mu_e-E_{\mathrm{int},kg}+\mu_g)/\hbar-\tilde{\delta}= (E_{\mathrm{int},ke}-\mu_e-E_{int,kg}+\mu_g)/\hbar
-\delta-(\mu_g-\mu_e)/\hbar=
(E_{\mathrm{int},ke}-E_{\mathrm{int},kg})/\hbar-\delta$. 

In the case where one state ($g$) is of BCS type and the final state $e$ is non-interacting
(and thus in normal state) the current becomes, from (\ref{ManyBCurrTorma})
\begin{eqnarray}
I(\tilde{\delta}) 
&=& \frac{\pi}{2} \sum_{k} \left| \Omega \right|^2  
(-u_k^2 n_F(E_{ke})n_F(E_{kg}) \delta((E_{ke}+E_{kg})/\hbar-\tilde{\delta}) \nonumber \\
&&-v_k^2 n_F(E_{ke})(1 - n_F(E_{kg})) \delta((E_{ke}-E_{kg})/\hbar-\tilde{\delta}) \nonumber \\
&&+u_k^2 (1-n_F(E_{ke}))(1-n_F(E_{kg})) \delta((E_{ke}+E_{kg})/\hbar-\tilde{\delta}) \nonumber \\
&&+v_k^2 (1-n_F(E_{ke}))n_F(E_{kg}) \delta((E_{ke}-E_{kg})/\hbar-\tilde{\delta}))  ,  \label{RFManyBCurrTorma}
\end{eqnarray}          
using the BCS quasiparticle energy $E_{kg} = \sqrt{\xi_{kg}^2 + \Delta^2}$, where $\Delta$ is the superfluid order 
parameter (the excitation gap) and $\xi_{kg}=\epsilon_{kg}-\mu_g$, where $\epsilon_{kg}$ is the kinetic energy. 

\subsubsection{Momentum-resolved RF spectroscopy (photoemission spectroscopy)}
 
In case of momentum resolved spectroscopy, one removes the $k$-summation from the above formulas and expresses the
spectrum as a function of $k$: $I(\tilde{\delta},k)$. 
The important feature of the momentum-resolved RF spectroscopy is that it 
can give the spectral function: one can understand the peak of $I(\tilde{\delta},k)\equiv I(E,k)$ as the dispersion relation $E(k)$.
Momentum-resolved RF spectroscopy is quite directly linked to the spectral function $A(k,\omega)$. 
Using Green's functions, the spectral function is the imaginary part of the retarded Green's function: $A(k,\omega)=-2 {\mathcal Im} (G_{ret}(k,\omega))$. 
The final result for momentum-resolved RF spectroscopy is, given in terms of
spectral functions as derived in Refs.~\cite{Torma2000} and \cite{Bruun2001} (a factor of $1/4$ is added to correspond to the definition
of field-matter interaction used in this article, c.f.\ Eq.~(\ref{EqHPtorma})):
\begin{eqnarray}
I(\tilde{\delta}, k) 
&=& - \frac{1}{4} \sum_{l} \left| \int d^3r' \Omega\left(\mathbf{r}\right)  
\varphi^*_{ke}({\bf r})\varphi_{lg}({\bf r'}) \right|^2   \nonumber \\
&&\int_{-\infty}^\infty \frac{d\epsilon}{2\pi} [n_F(\epsilon)-n_F(\epsilon - \tilde{\delta})] A_e(k,\epsilon) A_g(l,\epsilon - \tilde{\delta}) \nonumber \\
&=& -  \frac{|\Omega|^2}{4}  [n_F(\xi_k)-n_F(\xi_k - \tilde{\delta})] A_g(k,\xi_k - \tilde{\delta}) , \label{SpectralFunctionNewTorma}
\end{eqnarray}
where in the second line we have used the fact that for the RF field $\Omega\left(\mathbf{r}\right)\simeq \Omega$ (within the scale of the trap)
imposing momentum conservation, and taken the final state $e$ to be in a normal state for which the spectral function is straightforward:
$A_e(k,\epsilon)= 2 \pi \delta(\epsilon - \xi_{ke})$. In this case, the momentum-resolved RF 
spectrum directly provides 
the spectral function of the state of interest, namely the initial state.  

\subsubsection{Imbalanced gases}

The general formalism presented above applies directly to imbalanced
gases as well: one just needs to introduce, for instance, a bit more complicated set of Bogoliubov eigenenergies. For further information on the details of the Bogoliubov transformation for imbalanced gases, describing a superfluid, see for instance Ref.~\cite{Bakhtiari2008}. For the probing of the normal state for imbalanced gases see for instance Ref.~\cite{Veillette2008}. For further theory literature on the topic of imbalanced gases see the reviews by Radzihovsky and Sheehy \cite{Radzihovsky2010}, Chevy and Mora \cite{Chevy2010} and Gubbels and Stoof \cite{Gubbels2013}.  

\subsection{Bragg spectroscopy} 

Bragg spectroscopy does not change the 
internal state of the
particle and it can give a finite amount of momentum to the particle. In Bragg 
spectroscopy the initial state
corresponds to $\hat{c}^\dagger_{kg}$ and the final one to $\hat{c}^\dagger_{k+q,g}$ where $q$ is the momentum given to the atoms by the field. 
Typically,
two (nearly) counterpropagating laser beams are used: the atom is excited by absorbing one photon from 
one of the beams and
then goes back to the initial internal state by emitting a photon to the other beam. In the absence of line shifts due to 
interactions etc., the Bragg process is resonant at the frequency difference 
$2 \hbar k_L^2/m$ between
the beams, which corresponds to the recoil energy. The detuning $\delta$ can be tuned to be larger than this in order to create excitations in an interacting gas. 
Bragg spectroscopy can
probe density--density correlations and spin 
susceptibility. 

Linear 
response derivation for Bragg spectroscopy proceeds along the same basic principles as used in Section \ref{basicLinRespTorma}, 
but there are a few subtleties on the way. The label $e$ is
replaced by $g$ since the internal state does not change, except in a spin-flip process where 
$e$ is kept but now denotes the other component in the gas (say $g'$).
The intermediate state involved in the middle of the Bragg process is adiabatically eliminated: both 
the initial 
and final internal states are $g$, but momentum and energy are given. This means that the field coupling has 
spatial and temporal dependence of the form
$\Omega_\mathrm{eff}\left(\mathbf{r}\right) = \Omega_\mathrm{eff} \cos (\mathbf{q}\cdot\mathbf{r} - \delta t)$. Here $\Omega_\mathrm{eff}$ is the effective coupling related to the
two-photon transition, that is, it is proportional to $\Omega_1 \Omega_2$, the product of the Rabi frequencies related to the two
laser beams. 

The response is calculated by evaluating the density (or particle number) as a function of the momentum and energy given, $\mathbf{q}$ and $\delta$.   
The final result becomes (setting $t=0$ and $t_0=
-\infty$) 
\begin{equation}
N(\mathbf{q}, \delta) = \frac{\hbar\Omega_\mathrm{eff}}{2} \chi'' (\mathbf{q}, \delta) ,
\end{equation}
where $\chi''$ is the imaginary part of 
the density -- density susceptibility
\begin{equation}
\chi (\mathbf{k}, \omega) = - \frac{1}{\hbar Z} \sum_{m, n} e^{-\beta E_m} \left[ 
\frac{|\langle m | \hat{n}(\mathbf{k}) | n\rangle |^2}{\omega - \omega_{nm} + i\eta} - 
\frac{|\langle m | \hat{n}(-\mathbf{k}) | n\rangle|^2}{\omega + \omega_{nm} + i\eta} \right],  \label{chidefinitonTorma}
\end{equation}
and $Z$ is the partition function. The imaginary part of the susceptibility gives the dynamic
structure factor $S(\mathbf{q}, \delta)$: 
\begin{equation}
\chi''(\mathbf{q},\delta) = \pi (S(\mathbf{q}, \delta) - S(-\mathbf{q}, -\delta)) ,
\end{equation}
which becomes at zero temperature
\begin{equation}
\chi''(\mathbf{q},\delta) = \pi S(\mathbf{q}, \delta)
\end{equation}
and at finite temperature
\begin{equation}
\chi''(\mathbf{q},\delta) = \pi (1 - e^{-\hbar \delta/(k_B T)})S(\mathbf{q}, \delta) .
\end{equation}
Bragg spectroscopy therefore reveals the dynamic structure factor.
The dynamic structure factor incorporates all excitations
available in the system at frequency $\omega$ and momentum $\mathbf{k}$. In general, the dynamic structure factor contains both 
the density -- density
correlations and the spin susceptibility, that is, $S(\mathbf{k},\omega) = S(\mathbf{k},\omega)_{\uparrow \uparrow} 
+ S(\mathbf{k},\omega)_{\downarrow \downarrow}
+ S(\mathbf{k},\omega)_{\uparrow \downarrow}+S(\mathbf{k},\omega)_{\downarrow \uparrow}$. Experimentally, in case of a specific spectroscopy, one has to consider however whether
the perturbation couples to the density or the spin or both. Importantly, the Bragg response contains both single particle and collective excitations of the systems. For further information, see \cite{Torma2015}, Section 10.5.4. A derivation of the Bragg spectroscopy response by perturbation theory, 
using a slightly different approach from the one in \cite{Torma2015}, can be found in the book by Pitaevskii and Stringari~\cite{Pitaevskii2003}. 

\subsection{Lattice modulation spectroscopy} \label{latticemodulationTorma}

In the lattice modulation spectroscopy, the light potential that traps the particles into a lattice configuration oscillates with the frequency $\omega_L$ and excites the system via coupling to the density. 

Let us now consider a Hubbard type Hamiltonian in the lattice basis ($U$ is the on-site interaction energy and $J$ the hopping matrix
element, $u_{\sigma \sigma'}$ describes which types of interactions are present): 
\begin{eqnarray}
\hat{H} &=& - J \hat{H}_K + U \hat{H}_U + g(t)\hat{H}_K \\
\hat{H}_K &=& \sum_{\langle i, j, \rangle \sigma} \hat{c}^\dagger_{j\sigma} \hat{c}_{i\sigma} + h.c. \\
\hat{H}_U &=& \sum_{i,\sigma,\sigma'}  u_{\sigma \sigma'} \hat{n}_{i\sigma} \hat{n}_{i\sigma'} ,
\end{eqnarray}
where $g(t)$ describes the time modulation of the hopping.
For bosons, the relevant observable can be the heating caused by
the lattice modulation. In case of fermions, the observable of interest 
is the double occupancy $\hat{D}=\sum_{i} \hat{n}_{i\sigma} \hat{n}_{i\sigma'}$.
As this observable does not include the
field, one has to use the second order term Equation (\ref{General2ndFullCurrentTorma}) of the perturbation expansion. 
The double occupancy becomes (choose $t_0 = -\infty$)
\begin{equation}
\langle \Psi (t) | \hat{D} (t)  | \Psi (t) \rangle =  
\int_{-\infty}^\infty \int_{-\infty}^\infty dt' dt'' g(t')g(t'')\chi (t, t', t'') ,
\end{equation}
where 
\begin{eqnarray}
\chi (t, t', t'') = - \frac{\theta(t-t')\theta(t'-t'')}{\hbar^2}  \langle [[\hat{H}_U(t), \hat{H}_K(t') ], \hat{H}_K(t'')] \rangle.
\end{eqnarray}
Some terms from this expression average to zero
\cite{Kollath2006} so that in the end one is left with the non-oscillating term 
\begin{equation}
\langle \Psi (t) | \hat{D} (t)  | \Psi (t) \rangle = 
\langle \Psi (0) | \hat{D} (0)  | \Psi (0) \rangle - \frac{1}{2U} g(0)^2 \omega t {\mathcal Im} \chi_K(\omega),
\end{equation}
where $\chi_K(\omega)$ is the Fourier transform of $\chi(t) = (1/\hbar) \langle [H_K(t),H_K(0)]\rangle$.

\section{Early theory for spectroscopies of ultracold Fermi gases} \label{RFEarlyTheory}

\subsection{Spectroscopy for measuring the pairing gap}

In year 2000, it was suggested by T\"orm\"a and Zoller \cite{Torma2000} that the pairing gap in attractively 
interacting Fermi gases could be probed 
spectroscopically, and this reference gives the basic theory results for RF spectroscopy response in case of superfluids described by the
Bardeen--Cooper--Schrieffer (BCS) theory. Originally, Raman spectroscopy was suggested in \cite{Torma2000} but the concept and the theory description is are the same independent  of whether RF field or a Raman configuration is used. 

In experiments with trapped ultracold gases, the trapping 
geometry, often harmonic, may affect the results considerably. 
Ref.~\cite{Torma2000} gave results both for a homogeneous system and also for the trapped case where the basis states are
the harmonic oscillator states instead of the plane waves. The trapped case was considered in more depth by Bruun, T\"orm\"a, Rodr{\'i}gues and Zoller \cite{Bruun2001}, and also in later works by Ohashi and Griffin \cite{Ohashi20052} and He, Chen and Levin \cite{He2005}. One can also use the local
density approximation, that is, divide the system in spatial regions where the external 
potential is assumed to be locally constant and provide a simple shift in the chemical 
potential, $\mu_{\mathrm{eff}} = \mu - V_T(\mathbf{x})$. The homogeneous case 
spectra can be calculated in each region, and these can be averaged over the whole trap. Such an approach was used by Kinnunen, Rodr{\'i}gues and T\"orm\"a \cite{Kinnunen2004} and in many subsequent works.

The equations (\ref{ManyBCurrTorma}), (\ref{RFManyBCurrTorma}) and (\ref{SpectralFunctionNewTorma}) are given by Ref.~\cite{Torma2000}. They predict that in an experiment, in case of BCS pairing, the spectral peak that would in non-interaction case coincide with zero detuning $\delta = 0$ will shift. Also the shape of the peak changes since the density of states has a gap and corresponding strong peaks at the edges of the gapped area. Intuitively, one can think that the probing field has to provide extra energy to break a Cooper pair so that is it possible to transfer a particle to a third state that is not part of the superfluid state. Thus there is a threshold for the RF response.   

\subsection{RF threshold} \label{RFthreholdSectionTorma}

Choosing $\delta>0$
and assuming a constant density
of states $\rho$, as well as no particles initially in the non-interacting $e$ final state, at zero temperature 
one obtains \cite{Torma2000} from (\ref{RFManyBCurrTorma})
\begin{eqnarray}
I (\delta) = \frac{\pi\Omega^2}{2\hbar} \rho \theta(\delta^2 - \Delta^2/\hbar^2 + 2 \delta \mu/\hbar) \frac{\Delta^2}
{\delta^2} , \label{thresholdPTorma}
\end{eqnarray}
where $\theta$ is the Heaviside step function. 
Applying the actual density of states of the BCS state the result 
becomes \cite{Torma2000,Bruun2001} 
\begin{equation}
I (\delta) = \frac{\Omega^2Vm^{3/2}}{4\pi\hbar^{7/2}} \theta(\delta^2 - \Delta^2/\hbar^2 + 2 \delta \mu/\hbar) 
\frac{\Delta^2}{\delta^2} \sqrt{\frac{\delta^2-\Delta^2/\hbar^2}{\delta}+2\frac{\mu}{\hbar}}
 , \label{thresholdFormPTorma}
\end{equation}
where the system volume $V$ and the particle mass $m$ enter
(one can also use $k_F^3 = 3\pi^2 N/V$ where $N$ is 
the total particle number).
From Equations (\ref{thresholdPTorma}) and (\ref{thresholdFormPTorma}), one can observe three important points. 

First, the threshold for the RF response, as set by the theta-function, is \cite{Torma2000} 
\begin{equation}
\hbar \delta_{\mathrm{threshold}} = \sqrt{\mu^2 + \Delta^2} - \mu . \label{threshold2PTorma}
\end{equation}
This means that the threshold deviates from the gap energy in case the final state is empty initially, since the smallest energies for particle
transfer correspond to the particles in the lowest momentum states. Also a Hartree energy can be included in the chemical
potential and therefore affects the threshold. Note that approximately (Taylor expansion) $\hbar \delta_{\mathrm{threshold}} \sim \Delta^2/(2\mu) \sim \Delta^2/(2E_F)$ although one should be cautious with this since in unitary Fermi gases the gap is not extremely small compared to the chemical potential, and the chemical potential and the Fermi energy deviate from each other considerably.
In case the final state is occupied to the same Fermi level as the rest of the gas, then $\hbar \delta = \Delta$ is the threshold: this response gives the gap directly since only particles around the
Fermi level are transferred. This is analogous to measuring the gap in the superconductor-normal metal tunneling
experiments by Giaver \cite{Giaever1960}.   

Second, the high-energy tail of the spectrum can be obtained from Equation (\ref{thresholdFormPTorma}) by taking the large $\delta$ limit
where the square root term gives essentially $\sqrt{\delta}$. 
This combined with $\frac{\Delta^2}{\delta^2}$ leads to
\begin{equation}
I (\delta)_{\delta \rightarrow \infty} = \frac{\Omega^2V\sqrt{\hbar}}{4\pi \sqrt{m}} \frac{m^2 \Delta^2}{\hbar^4} \frac{1}{\delta^{3/2}} .
\label{RFdecayTorma}
\end{equation}
The tail of the RF spectrum $I(\omega)$ therefore decays as $\omega^{-3/2}$.

The third important point in Equation (\ref{thresholdPTorma}) (and (\ref{thresholdFormPTorma}) as well as (\ref{RFdecayTorma})) is that the current is directly proportional to the square of the pairing gap $\Delta^2$.  

Note that the decay of the tail is in contrast to the superconductor -- normal metal
experiments where the current simply grows after the threshold. This is explained by the difference that momentum is not conserved in the tunneling
(particle transfer) process, while RF spectroscopy preserves momentum.

\subsection{Early theory work on Bragg spectroscopy}

In the context of ultracold Fermi gases, some of the first studies 
considering the prospects of Bragg spectroscopy 
theoretically are by Minguzzi, Ferrari and Castin \cite{Minguzzi2001}, 
Rodr{\'i}guez and T\"orm\"a~\cite{Rodriguez2002}, B\"uchler, Zoller and Zwerger~\cite{Buchler2004}, Bruun and Baym \cite{Bruun2006}, Combescot, Giorgini and Stringari~\cite{Combescot2006}, and Challis, Ballagh and Gardiner~\cite{Challis2007}. 

\section{Early RF spectroscopy experiments: mean-field effects} \label{RFExperimentsTorma}

The pioneering experiments opening 
the route towards the use of RF spectroscopy in studies of fermionic many-body physics were the works of Regal and Jin \cite{Regal2003} and Gupta, Ketterle and coworkers \cite{Gupta2003} in the year 2003 (for the earlier use of
RF spectroscopy in probing condensates of bosonic atoms see the introduction in Ref.~\cite{Gupta2003}).
In Refs.~\cite{Regal2003} and \cite{Gupta2003}, RF spectroscopy was applied to study mean-field energies in strongly interacting Fermi gases. In Ref.~\cite{Regal2003} $^{40}$K atoms in two hyperfine states $|F=9/2,m_F=-9/2\rangle$ and $|9/2,-7/2\rangle$ were prepared. The atoms were basically non-interacting. Then an RF pulse was applied to move atoms from the state the $|9/2,-7/2\rangle$ to the state $|9/2,-5/2\rangle$ which was strongly interacting with atoms in state $|9/2,-9/2\rangle$ due to a Feshbach resonance. The resulting mean-field energy shift, see Equation (\ref{RFnormalTorma}) and the discussion following it, was observed. In Ref.~\cite{Gupta2003}, the system was prepared in a mixture of two lowest hyperfine states of $^{6}$Li, $|1/2,1/2\rangle\equiv |1\rangle$ and $|1/2,-1/2\rangle\equiv |2\rangle$ and an RF pulse was applied between the latter and a third state, $|3/2,-3/2\rangle\equiv |3\rangle$, to observe mean-field shifts in the energy (the shorthand notation $|1\rangle$, $|2\rangle$, $|3\rangle$ for $^{6}$Li will be used later in this article). Also another experiment, tranferring atoms initially in the state $^{6}$Li, $|1/2,1/2\rangle$ to the state $|1/2,-1/2\rangle$ with an RF pulse, was performed and resulted in a totally different outcome, as discussed in the following section.

\subsection{RF spectroscopy reveals symmetries of the Hamiltonian} \label{RFHamiltonianSymmetries}

In ultracold gas spectroscopies, the process can often be highly coherent. When applying a field that couples atoms in an internal state $g$ to another one, say $e$, for a certain time duration $\tau$, the result is that all atoms in the system are in a coherent superposition $a(\tau)|g\rangle+b(\tau)|e\rangle$ of the two internal states. 

If the field causes basically a rotation in the space spanned by the degrees of freedom of the system, for instance the (pseudo)spins (internal states), one has to check whether the Hamiltonian of the system has related symmetries. In other words, is it  invariant under such a rotation? An intriguing experimental example of this is presented by Gupta, Hadzibabic, Zwierlein and Ketterle, first in \cite{Gupta2003}, and then in  \cite{Zwierlein2003} which also contains a theoretical analysis of the phenomenon. 
In Ref.~\cite{Zwierlein2003} a two-component Fermi gas is considered, let us label the components by $g$ and $e$. 
In case of contact interaction and no trap potential, the many-body Hamiltonian of the system, with no field, is 
(see also Equation (\ref{EqHPtorma}))
\begin{eqnarray}
\hat{H}_C&=&\int d^3r \sum_{\sigma=e,g}\hat{\psi}_{\sigma}^{\dagger}\left(\mathbf{r}\right)\left(-\frac{\hbar^{2}\nabla^{2}}
{2m_{\sigma}}
+E_{\sigma}\right)\hat{\psi}_{\sigma}\left(\mathbf{r}\right) \nonumber \\ 
&+& V \int d^3r \hat{\psi}_{e}^{\dagger}\left(\mathbf{r}\right)\hat{\psi}_{g}^{\dagger}\left(\mathbf{r}\right)\hat{\psi}_{g}\left(\mathbf{r}\right)\hat{\psi}_{e}\left(\mathbf{r}\right). 
 \label{EqRotationMFtorma}
\end{eqnarray} 
Assuming a mean-field theory that includes only the Hartree energy, the interaction 
term of the last line of Equation (\ref{EqRotationMFtorma}) 
gives, say, for component $g$, simply $E_{int}= V n_e \hat{\psi}_{g}^{\dagger}
\hat{\psi}_{g}$ where $n_e = \langle \hat{\psi}_{e}^{\dagger}
\hat{\psi}_{e}\rangle$. If all atoms are initially in state $g$, then $n_e=0$ implies $E_{int}=0$. 
Now think that one particle would be transferred to the state $e$: clearly $E_{int}$ is finite. Following such argument, one should observe an interaction energy shift in the frequency required to transfer the particle from one internal state to another since the corresponding mean-field energies are different: the resonance condition is not $\delta=0$ but $\delta-E_{int}=0$. However, the case where the gas is probed by a coherent field actually produces a completely different result. This can be seen by noting that the coherent rotation caused by the field can be expressed in the generic form
\begin{eqnarray}
\hat{\psi}_{1} &=& \cos (\theta/2) e^{\phi/2} \hat{\psi}_{g} + \sin (\theta/2) e^{-\phi/2} \hat{\psi}_{e} \\
\hat{\psi}_{2} &=& - \sin (\theta/2) e^{\phi/2} \hat{\psi}_{g} + \cos (\theta/2) e^{-\phi/2} \hat{\psi}_{e} . \label{RotatedComponentstorma}
\end{eqnarray}
Here $\theta$ and $\phi$ define the rotation.
Applying this transformation to the Hamiltonian (\ref{EqRotationMFtorma}) actually keeps the Hamiltonian unchanged. This was indeed observed in the experiments: no
interaction energy shift was detected in the RF transfer between the states $|1\rangle$ and $|2\rangle$ in \cite{Gupta2003}. In the further study \cite{Zwierlein2003}, the same group connected the absence of the mean-field energy shift to the symmetries of the Hamiltonian and thereby showed that no shifts should be observed even in the case of a paired superfluid. Similar phenomena have been observed in NMR experiments of $^{3}$He superfluids by Osheroff, Gully, Richardson, and Lee 
\cite{Osheroff1972}. There, in an NMR experiment a coherent superposition of two spin states is created in a similar way as in the RF spectroscopy experiments here. If the Hamiltonian is symmetric with respect to the rotation, no shifts are expected. The appearance of a shift for in \cite{Osheroff1972} in certain parameter regimes was suggested by Leggett \cite{Leggett1972} to be related to spontaneous breaking of the symmetry in question, such as an unisotropic BCS-type superfluid with pairing in a higher angular momentum state. Note that in a typical ultracold gas Fermi superfluid, any shifts should be absent, as mentioned above. This is because even when the superfluid has a spontaneously broken symmetry in the global phase, it is still symmetric with respect to a coherent rotation of the components, see above. In a usual s-wave BCS superfluid, one can rotate the spins to a new basis, and the physics will be the same. This is, however, not necessarily the case for more complicated superfluids, such as the $^{3}$He example.  

\section{RF spectroscopy in understanding the BCS-BEC crossover: experiments and theory using the density balanced gas}

\subsection{The pioneering experiments}

Fermionic many-body pairing and superfluidity in ultracold gases was gradually established during 2003-2005 by contributions from several groups. Insprired by predictions of superfluidity in ultracold Fermi gases by Houbiers, Stoof, Hulet and coworkers \cite{Houbiers1997}, and by the breakthrough experiment reaching Fermi degeneracy of potassium atoms by DeMarco and Jin \cite{DeMarco1999}, the field had set going in the late 1990's. One great achievement of the ultracold Fermi gas research so far is the experimental verification of the BCS-BEC crossover. Already in the 1960's and 1980's it was suggested by Keldysh (see \cite{Keldysh1995}), Eagles \cite{Eagles1969}, Leggett \cite{Leggett1980}, and Nozieres and Schmitt-Rink \cite{Nozieres1985} that the apparently different phenomena of superconductivity as condensation of delocalized Cooper pairs and Bose-Einstein condensation of tightly bound, localized pairs (such as molecules) could actually be connected to each other via a smooth crossover. Leggett showed that the simple wavefunctions describing both ends, $ |\Psi\rangle_{BEC}$ and 
$|\Psi\rangle_{BCS}$, are actually smoothly connected to each other, that is
\begin{equation} \label{eq:BCS_PT}
 |\Psi\rangle_{BEC} = \mathcal{N} e^{\lambda \sum_{\mathbf{k}} \varphi_\mathbf{k} \hat c^\dag_{\mathbf{k}\uparrow} \hat c_{-\mathbf{k}\downarrow}^\dag} | 0 \rangle
=
\prod_{\mathbf{k}} (u_\mathbf{k} + v_\mathbf{k}
 \hat c^\dag_{\mathbf{k}\uparrow} \hat c_{-\mathbf{k}\downarrow}^\dag) | 0 \rangle 
= | \Psi\rangle_{BCS} ,
\end{equation}
if one requires $v_\mathbf{k} / u_\mathbf{k} = \lambda\varphi_\mathbf{k}$ and $\mathcal{N}
= \prod_\mathbf{k} u_\mathbf{k}$. Here $\varphi_\mathbf{k}$ describe the internal structure of the composite boson, $|\lambda|$ divided by the volume gives the condensate fraction, and $v_\mathbf{k}$, $u_\mathbf{k}$ are the BCS coherence factors (for more details see, e.g., Ref.~\cite{Ketterle2008} or Ref.~\cite{Parish2015}).
 
However, these arguments were based on mean-field theory, with approximate ways of including fluctuations; exact calculations of the intermediate regime between the BCS and BEC regimes, at finite temperature, are exceedingly difficult. Experimentally it was difficult to find a system where one could tune the parameters in a controlled and continuous manner: the various types of superconductors and superfluids discovered were naturally in a certain place in the crossover and dramatic tuning of the interaction was not possible \cite{Randeria2012}. Therefore it remained as an open question whether the transition from one regime to the other is a smooth crossover or whether, for instance, quantum phase transitions occur. It was only when the ultracold Fermi gas  experiments arrived that the existence of the crossover could be confirmed experimentally. 

Basically, the achievement of fermionic pairing and superfluidity went hand in hand with the studies of the BCS-BEC crossover in ultracold Fermi gases. Molecule formation at the BEC side of the crossover was first observed by the groups of Jin, Hulet, Salomon and Grimm, see Regal {\it et al.}~\cite{Regal2003b}, Strecker {\it et al.}~\cite{Strecker2003}, Cubizolles {\it et al.}~\cite{Cubizolles2003} and Jochim {\it et al.}~\cite{Jochim2003}. Such molecules were found to form a Bose-Einstein condensate by the Jochim, Grimm and coworkers \cite{Jochim2003b}, Regal, Greiner and Jin \cite{Regal2003c} and Zwierlein, Ketterle and coworkers \cite{Zwierlein2003zero}. Condensation of pairs at the unitarity regime between the BCS and BEC sides was achieved by Regal, Greiner and Jin \cite{Regal2004} and Zwierlein, Ketterle and coworkers \cite{Zwierlein2004}. The whole crossover was studied in various experiments showing, among other things, continuous development of the critical temperature, collective modes, and pairing energies over the crossover, by Bartenstein, Grimm and coworkers \cite{Bartenstein2004}, Kinast, Thomas and coworkers \cite{Kinast2004} and Bourdel, Salomon and coworkers \cite{Bourdel2004}. The smoking gun proof of superfluidity of the gas was provided by the experiment by Zwierlein, Ketterle and coworkers \cite{Zwierlein2005,Zwierlein2006} where quantized vorticies were observed throughout the crossover. 
 
\begin{figure}
\begin{minipage}[c]{0.45\textwidth}
\includegraphics[width=\textwidth]{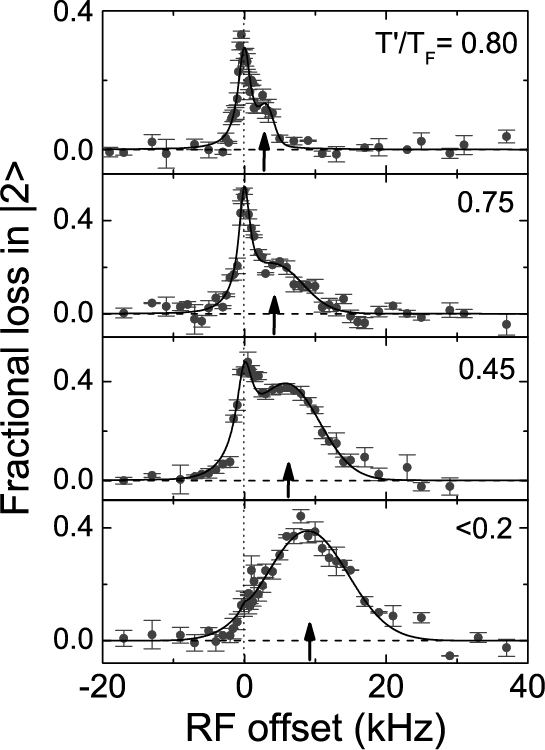}
\end{minipage}
\begin{minipage}[c]{0.6\textwidth}
\includegraphics[width=\textwidth]{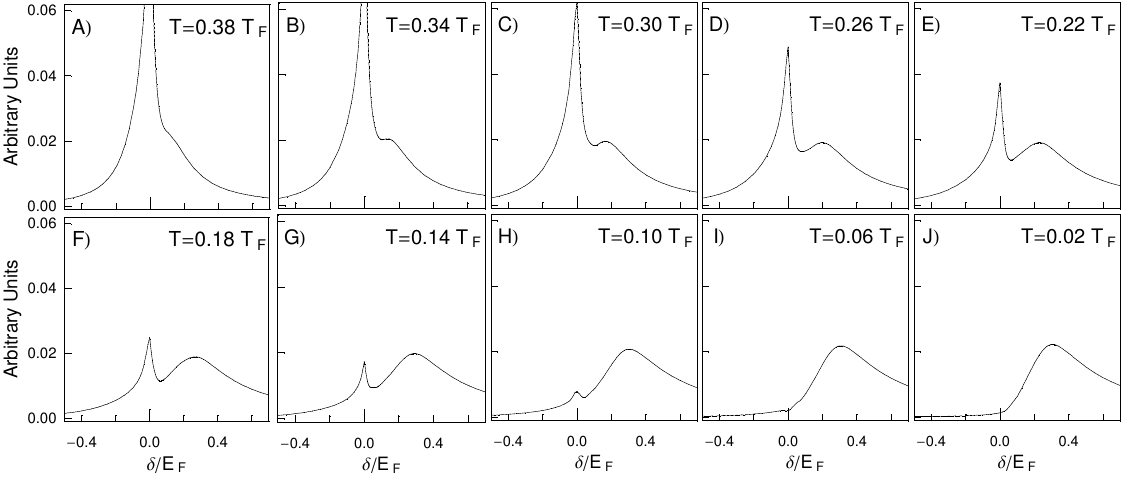}
\end{minipage}
\caption{Above: RF spectroscopy was used in \cite{Chin2004} to study many-body pairing of fermions. The shift of the peak away from zero RF offset (detuning) reflects pairing which emerges when temperature is lowered. Below: Theoretical modeling of the spectra in \cite{Kinnunen2004}. The double peak structure originates from paired particles in the middle of the trap and unpaired ones (the zero detuning $\delta=0$ peak) at the trap edges where the local chemical potential is lower. Here $T_F$ is the Fermi temperature.} 
\label{PairingGapImagesTorma}
\end{figure}

Here, I will say a few more words about a work which was one of the crucial experiments establishing the crossover and where RF spectroscopy was used. In 2004, RF spectroscopy revealed signatures of fermionic many-body pairing in a strongly interacting Fermi gas in an experiment by Chin, Grimm, and coworkers \cite{Chin2004}. As discussed above (Equations (\ref{RFManyBCurrTorma}), (\ref{thresholdPTorma})--(\ref{threshold2PTorma})), existence of a pairing gap should lead to a shift of the RF response with respect to the non-interacting case. Indeed a shift in the response was observed when the gas was cooled down, see Figure \ref{PairingGapImagesTorma}. It was also observed that the shift was proportional to the Fermi energy: this showed that non-trivial fermion pairing was taking place instead of only molecule formation, where the binding energy also leads to shifts in RF spectra \cite{Chin2005,Bartenstein2005} but with no significant dependence on density. Interestingly, the spectra at higher temperatures displayed a two-peak structure as visible in Figure \ref{PairingGapImagesTorma}. In an accompanying work, Kinnunen, Rodr{\'i}guez and T\"orm\"a \cite{Kinnunen2004} provided a theoretical framework which supported the conclusion that the two peaks originate from paired particles in the middle of the harmonic trap and unpaired ones at the edges. The density of particles decreases in a harmonic trap from the center towards the edges, and with it also the critical temperature related to many-body pairing. In the lowest temperatures, only one peak was observed. Other theory work following the experiment by He, Chen and Levin \cite{He2005}, and Ohashi and Griffin \cite{Ohashi2005,Ohashi20052} reached similar conclusions.

The standard RF spectroscopy basically gives the spectral function $A(\mathbf{k},\omega)$ integrated over the momentum $\mathbf{k}$. Therefore the experiment \cite{Chin2004} was somewhat analogous to the first tunneling spectroscopy measurements by Giaver that revealed the pairing gap in superconductors \cite{Giaever1960}. However, the RF response has a 
spectral shape different from the IV-curve due to momentum conservation 
in the spectroscopy \cite{Torma2000}; the IV curve has an onset at the gap energy and thereafter simply grows, while as seen in Figure \ref{PairingGapImagesTorma}, the RF response has a decaying tail. This tail, exactly due to the momentum conservation, actually contains interesting information, as will be discussed in Section \ref{contactTorma} of this article.

Although trapping effects may be considered even helpful in analysing the RF spectroscopy experiments, such as the case of the two peaks discussed above, it is desirable to be able to probe also the bulk properties of the gas without trap averaging. Shin, Ketterle and coworkers developed in 2007 a tomographic version of RF spectroscopy \cite{Shin2007} which is very powerful since it is spatially resolved: different spatial locations of the trap correspond to different densities, thereby one receives information, for instance about pairing, that corresponds to bulk systems with different chemical potentials. This thinking, of course, assumes that the local density approximation is valid. In 2012, Sagi, Jin and coworkers developed another method to avoid trap averaging in RF spectroscopy \cite{Sagi2012}: they intersect
two perpendicularly propagating hollow light beams
that optically pump atoms at the edge of the cloud into a
spin state that is dark to the detection.

\subsection{Exploring fermion pairing by RF spectroscopy}

In section \ref{simpleLinearResponseTorma}, it was discussed that the simple factorization of the four-point correlator in Equation (\ref{FactorizedCurrentTorma}) is accurate only if the final state in the spectroscopy does not significantly interact with the rest of the system. Sometimes this is, however, not the case in ultracold gases. For instance in case of  $^6$Li, if one uses in RF spectroscopy the hyperfine states 
$|1\rangle$ ($|F=1/2,m_F=1/2\rangle$) and $|2\rangle$ ($|F=1/2,m_F=-1/2\rangle$) as the ones that are strongly interacting or paired, 
and as the final state the state $|3\rangle$ ($|F=3/2,m_F=-3/2\rangle$) (as is the case
in Refs.~\cite{Gupta2003} and \cite{Chin2004} and several subsequent works), then
final state interactions cannot be neglected. 

It was pointed out by Yu and Baym that final state interactions should be taken into account in extracting the value of the pairing gap from RF data \cite{Yu2006}. This can be done with the help of elaborate theoretical analysis as in the theory work of Perali, Pieri and Strinati \cite{Perali2008,Pieri2009}, and the work combining experiment and theory by the Grimm and Strinati groups \cite{Pieri2011}. The other option is to avoid final-state interactions altogether. For $^{40}$K, final state interactions can be neglected because the Feshbach resonances are narrow and there are no overlaps between different resonances. Concerning $^{6}$Li, Schunck, Ketterle and coworkers discovered that if one uses 
$|1\rangle$ and $|3\rangle$ as the initial states and $|2\rangle$ as the final one, the final state interactions become 
negligible \cite{Schunck2008}. This is because $^6$Li has broad resonances and the ones between $|1\rangle$ and $|2\rangle$ as well as 
$|2\rangle$ and $|3\rangle$ are partly overlapping while the $|1\rangle$ and $|3\rangle$ resonance is well separated
in magnetic field from the others. They used RF spectroscopy in this new setting to determine the fermion pair size. Knowing $\mu$, one can obtain the gap $\Delta$ from observing the threshold, Equation (\ref{threshold2PTorma}).
Such a threshold and RF line shapes of the form of Equations (\ref{thresholdPTorma}) and (\ref{thresholdFormPTorma}) were observed in Ref.~\cite{Schirotzek2008}, as will be discussed in Section \ref{QuasiparticleSpectroscopy}. 

Sum rules can be used for finding the average peak position in RF spectroscopy.
This was done by Yu and Baym \cite{Yu2006} (see also the works of Punk and Zwerger \cite{Punk2007}, Baym, Pethick, Yu and Zwierlein \cite{Baym2007}, and Basu and Mueller \cite{Basu2008}). When there are strong final state interactions, the spectrum is expected to be quite symmetric, and the 
result becomes (in the notation of Ref.~\cite{Yu2006})
\begin{equation}
\delta_{\mathrm{peak}} = (g_{eg}-g_{gg'}) \frac{\Delta^2}{n_0 g_{eg} g_{gg'}} ,   \label{YuSumRuleTorma}
\end{equation}
where $g_{eg}$ is the interaction coupling between the initial and final states and $g_{gg'}$ marks the coupling responsible
for the BCS pairing between the initial state and the other component of the gas ($g'$), and $n_0$ is the density. From Equation (\ref{YuSumRuleTorma}) one can see that if the symmetry of the Hamiltonian is preserved under the perturbation,
here the case is $g_{eg}=g_{gg'}$, there is no shift; this is consistent with the 
discussion in Section \ref{RFHamiltonianSymmetries}. Furthermore, the peak position is proportional to $\Delta^2$. If there are no 
final state interactions ($g_{eg}=0$), the derivation in Sections \ref{RFTorma} and \ref{RFthreholdSectionTorma}, following to the early work by T\"orm\"a and coworkers \cite{Torma2000,Bruun2001}, gives the
threshold of Equation (\ref{threshold2PTorma}), namely $\hbar \delta_{\mathrm{threshold}} 
= \sqrt{\mu^2 + \Delta^2} - \mu \simeq \frac{\Delta^2}{2\mu}$.
Therefore, both in cases of finite and no final state interactions there is a $\Delta^2$ dependence,
although there is no direct comparison since (\ref{YuSumRuleTorma}) is not valid for
$g_{eg}=0$. 

In case of no final-state interactions, $g_{eg}=0$, the spectrum has an asymmetric shape with a long, slowly decaying tail. Sum-rule given mean values of the response then become cut-off-dependent within
the BCS theory \cite{Yu2006}. Thus for such shapes a mean value of the frequency is not a useful characterization and
other approaches are preferred over sum rules, such as those discussed in Sections \ref{RFTorma} and \ref{RFthreholdSectionTorma}.  

\subsection{Momentum-resolved RF spectroscopy (photoemission spectroscopy)}
A momentum-resolved version of RF spectroscopy was introduced by Stewart, Gaebler and Jin \cite{Stewart2008}. 
Avoiding the momentum integration, the spectral
function $A(\mathbf{k},\omega)$ was directly observed in the experiment. In this sense, momentum-resolved RF spectroscopy is analogous to angle-resolved photo-emission spectroscopy (ARPES); for further information see for instance \cite{Chen2009,Georges2012}

\subsection{Beyond BCS theory}

All the analytical results in Sections \ref{RFTorma} and \ref{RFthreholdSectionTorma} in this article are based on the BCS mean-field theory. Although this is instrutive to understand what kind of qualitative behaviour might be expected from spectroscopies, strongly interacting ultracold Fermi gases are obviously asking for beyond mean-field theoretical treatments. 
Simple mean-field theories are based on having well-defined quasiparticles and 
this assumption might not be valid in strongly interacting systems. Even if yes, the properties of the quasiparticles 
may have to be evaluated beyond BCS theory. The BCS--BEC crossover at low temperature is surprisingly well
qualitatively described by the BCS--Leggett theory. But at higher temperatures the quantitative agreement is not good. 

Theoretical calculations of RF spectra, with 
the correlators evaluated with various many-body theories and approaches that go beyond the simple BCS theory are presented for instance in
Refs.~\cite{Kinnunen2004a,Kinnunen2004,He2005,Perali2008,Magierski2009,Haussmann2009} and in other works by 
these and several other groups. In particular, it was noted by Kinnunen, Rodr{\'i}gues and T\"orm\"a \cite{Kinnunen2004a,Kinnunen2004} and by He, Chen and Levin \cite{He2005} that since pairing features may appear also above $T_c$ in beyond-BCS scenarios, the shift of the RF spectral peak may appear already in the normal state and is thus not alone a signature of superfluidity.

\section{Spin-imbalanced Fermi gases}

Although the topics discussed from here on are conceptually very related to the ones above, I separate them to a dedicated section because imposing a spin-density imbalance in a Fermi gas has spun off such an enourmous amount of interesting physics and perspectives, many still unexplored. In ultracold gases, it is possible to prepare an initial condition with different amounts of particles in different internal states (species). In ultracold Fermi gases, such studies were pioneered by experiments published in 2006 by Zwierlein, Shin, Ketterle and coworkers \cite{Zwierlein2006,Shin2006}, and by Partridge, Hulet and coworkers \cite{Partridge2006}. In \cite{Zwierlein2006}, the stability of vortices was studied as a function of polarization $P=(n_\uparrow - n_\downarrow)/(n_\uparrow + n_\downarrow)$ where $n_\sigma$ are the densities of the species, that is, the spin-density imbalance, on both sides of the BCS-BEC crossover. The famous Chandrasekhar-Clogston limit \cite{Chandrasekhar1962,Clogston1962} of chemical potential imbalance that destroys fermionic superfluidity was observed, leading to a critical polarization of about 0.8 in a harmonic trap in the works \cite{Zwierlein2006,Shin2006}. Similar critical polarization was later obtained in the experiments by Nascimb{\`e}ne, Salomon and coworkers \cite{Nascimbene2009}, and was also confirmed with quantum Monte Carlo calculations by Lobo, Recati, Giorgini and Stringari \cite{Lobo2006}, and by a dynamical mean-field theory method by Kim, Kinnunen, Martikainen and T\"orm\"a \cite{Kim2011}. In Ref.~\cite{Partridge2006}, a critical polarization close to one was reported, but it was later found to be due to depolarization in the evaporative cooling process \cite{Parish2009,Liao2011}. 

The work by Shin, Ketterle and coworkers in 2008 \cite{Shin2008} established by tomographic imaging of the densities that, in a harmonic trap, there is a balanced superfluid in the middle, then a sharp change, reflecting a first order phase transition, to a partially polarized normal state. Finally, at the very edges of the trap, the gas is fully polarized. The normal gas in the edges allows very accurate determination of temperature from cloud profiles. Utilizing this, Shin {\it et al.}\  determined the critical temperature of a balanced unitary Fermi gas to be $T_c/T_F \simeq 0.15$, where $T_F=(\hbar^2/2mk_B)(6\pi^2n_\sigma)^{2/3}$ is the Fermi temperature of the gas, $m$ the mass, and $n_\sigma$ the density of one species. This is very close to values given by various theoretical approaches and later experiments.   

The spin-imbalanced gases offered new interesting prospects for the use of spectroscopies as well, in particular precision measurements of the quasiparticle energies and the pairing gap and the studies of polarons, which will be discussed below.  

\section{Quasiparticle spectroscopy utilizing the spin-imbalance} \label{QuasiparticleSpectroscopy}

\begin{figure}
	\includegraphics[width=0.8\textwidth]{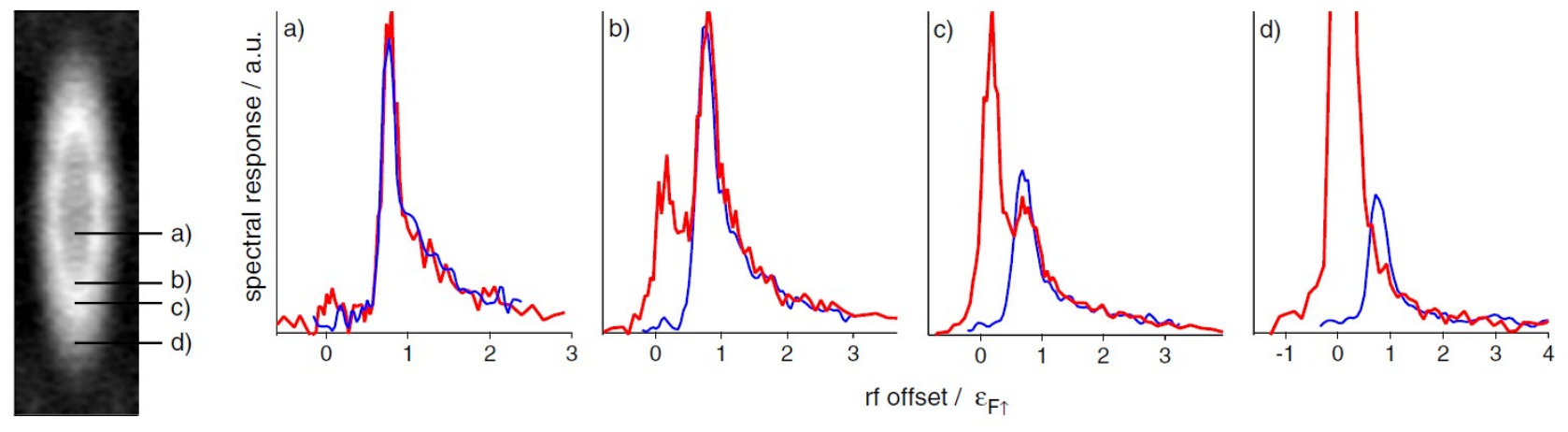}
	\caption{a) RF spectra from \cite{Schirotzek2008} for increasing local polarizations ranging from near zero in a) and 0.64 in d). The red curve is the majority spectrum and the blue the minority. Different polarizations are naturally obtained by considering spectra from different areas in the trapped gas, see the leftmost picture. In panels b) and c), the majority component shows both pairing and quasiparticles; from such data, values of the pairing gap and Hartree energy can be extracted. In d), the majority component does not show a peak feature under the minority peak which indicates that polaron-type physics takes place instead of pairing.} 
	\label{QuasiparticleImagesTorma}
\end{figure}

An important work where spectrocopies were heavily utilized is the one by Schirotzek, Ketterle and coworkers \cite{Schirotzek2008}. There tomographic RF spectroscopy was used in different areas of the trap: the fully balanced middle part and various polarized areas around it. Majority and minority spectra were separately analysed: in the balanced case, both spectra overlap and show a shift due to pairing, while in the imbalanced case unpaired majority particles show a contribution with no shift. Intriguinly, the chemical potential imbalance induces quasiparticles in the middle superfluid region already at a low temperature \cite{Parish2007}. Therefore Schirotzerk {\it et al.}\ were able to measure accurately the quasiparticle energies by the RF spectroscopy, see Figure \ref{QuasiparticleImagesTorma}. According to the predictions by T\"orm\"a, Zoller, Bruun and Rodr{\'i}guez \cite{Torma2000,Bruun2001}, the quasiparticle energies $E_k = \sqrt{\Delta^2 + (\epsilon_k + U - \mu)^2}$, where $U$ is the Hartree energy, appear in the RF spectra according to Equations (\ref{RFManyBCurrTorma}), (\ref{thresholdPTorma})--(\ref{threshold2PTorma}). This allows extracting the pairing gap and the Hartree shift in high precision from the RF spectra. Schirotzek {\it et al.}\ obtained $\Delta = 0.44(3)\epsilon_{F\uparrow}$ and $U=-0.43(3)\epsilon_{F\uparrow}$ which are in excellent agreement with the value $\Delta \simeq 0.46\epsilon_{F\uparrow}$ predicted by a Luttinger-Ward approach by Haussmann, Zwerger and coworkers \cite{Haussmann2007} and by quantum Monte Carlo calculations by Carlson and Sanjay \cite{Carlson2008}. By studying the extremely polarized gas, the work \cite{Schirotzek2008} contributed to the start of polaron physics studies in ultracold Fermi gases.    

\section{Spectroscopies in polaron studies}

The research on imbalanced Fermi gases naturally led to polaron studies. Thinking in terms of a single impurity immersed in a system, in order to reveal information about the ground state or even excitations, has been a very powerful approach in physics. Sometimes the behaviour of the impurity can be effectively described by the concept of a polaron: an object that deforms its surroundings slightly, causing for instance excitations, and thereby obtains an effective mass different from the bare mass of the impurity. A polaron fulfils the Fermi liquid paradigm and is associated with a quasiparticle weight/residue. 

It turned out that at the limit of large polarization (large imbalance), the physics of the ultracold Fermi gas is well described by polaron physics. Throughout the BCS-BEC crossover, there is a non-trivial evolution to a bound molecule. We do not discuss here all the important theory and experiments on this topic but refer to excellent reviews
by Chevy and Mora \cite{Chevy2010} and by Massignan, Zaccanti and Bruun \cite{Massignan2014}. We only wish to point out those experiments where spectroscopies provided crucial insight to polaron physics. 

Schirotzek, Zwierlein and coworkers measured the polaron energies in \cite{Schirotzek2009} by using RF spectroscopy. The RF spectra of the minority species revealed not only the polaron energy, but also the quasiparticle weight or quasiparticle residue $Z$: it is basically given by the spectral weight under the narrow quasiparticle peak in comparison to a broad incoherent background. They also observed that, during the BCS-BEC crossover, the Fermi polaron transfers into a bound molecule and this is connected to a quantum phase transition from a Fermi gas to a coexistence of Bose and Fermi gases.   

In May 2012, two breakthrough works using RF spectroscopy to study polarons were published back-to-back in Nature, namely a study of 
$^{6}$Li--$^{40}$K mixture by Kohstall, Grimm and coworkers \cite{Kohstall2012}, and experiments on a 2D ultracold $^{40}$K gas by Koschorrek, K{\"o}hl and coworkers \cite{Koschorreck2012}.

Kohstall, Grimm and coworkers studied polaron energies with RF spectroscopy in a system where the majority atoms were 
$^{6}$Li but the minority was another, 
heavier, species namely $^{40}$K. 
RF spectroscopy is typically done, as discussed already many times in this article, in such a way that particles in a species that is strongly interacting with another species are transferred by the RF pulse to a third state. Kohstall {\it et al.}\ did it, however, the other way round: they started with an initial state of potassium that does not interact strongly with lithium, and then transferred the potassium atoms (which were only a small amount, corresponding to the polaron/impurity limit) to a state that has a Feshbach resonance and thus interacts strongly with the lithium internal state. For historical reasons, this is called the {\it inverse RF spectroscopy} (curiously, the very first RF spectroscopy experiment on Fermi gases by Regal and Jin \cite{Regal2003} also used the "inverse" version, but since then all the works used the "standard" RF spectroscopy until the work of Kohstall {\it et al.}).
One remarkable aspect of the inverse RF spectroscopy is that it can easily probe excitations of the system: while in the usual version one typically probes properties of the ground state, and of the excitations present due to the (low) temperature, with the inverse process it is easy to provide the right amount of energy by the RF field to create a specific excitation of the interacting system. Kohstall {\it et al.}\ studied the whole crossover from attractive to repulsive polarons, and observed the polaron branches as well as a molecule-hole continuum.

The work by Koschorrek, K\"ohl and coworkers \cite{Koschorreck2012} also explored both the repulsive and attractive polarons. They used momentum-resolved RF spectroscopy (momentum-resolved photo-emission spectroscopy), which directly gives the single particle spectral functions, to study how the polaron energy behaves in two dimensions. Note that in two dimensions, a bound state (molecule) exists throughout the crossover and competes with the polaron branches. The repulsive polaron branch decay times were determined. 

\section{Two dimensional gases: studies of polarons and interaction energies by spectroscopies}

RF spectroscopy was used for measuring interaction energies and confinement induced resonances in a two-dimensional Fermi gas by Fr\"ohlich, K\"ohl and coworkers \cite{Frohlich2011}. Sommer, Zwierlein and coworkers studied the interaction and binding energies over the 2D-3D crossover \cite{Sommer2012}. 
RF spectroscopy was also used for studying quasi-2D Fermi gases by Zhang, Thomas and coworkers \cite{Zhang2012}. They discovered polaron states also in the quasi-2D setting. More about all these, and other work on 2D gases which did not use spectroscopies, can be read from the review by Levinsen and Parish \cite{Levinsen2015} and from, for instance, the recent work by Murthy, Jochim and coworkers \cite{Murthy2015} observing the Berezinskii-Kosterlitz-Thouless transition in a two-dimensinal Fermi gas.  

\section{Static and dynamic structure factors by Bragg spectroscopy}

Bragg scattering and spectroscopy have been used as a tool in ultracold gases research since the late 1980s.  Bose--Einstein condensates were for the first time studied by Bragg spectroscopy in Refs.~\cite{Kozuma1999} and \cite{Stenger1999}.
Bragg spectroscopy provides not only the single particle but 
also the collective mode spectrum of the gas. This is important concerning both bosonic and fermionic superfluids, and any
other many-body states. The
dynamic $S(\mathbf{k},\omega)$ and static $S(\mathbf{k})$ structure factors for Fermi gases were measured with this spectroscopy for the first time 
by Veeravalli, Vale and coworkers \cite{Veeravalli2008}.

\section{RF, hopping modulation and Bragg spectroscopies in lattices}

The use of RF spectroscopy for fermions in the context of
optical lattices was pioneered by the group of Esslinger in Moritz 
{\it et al.}~\cite{Moritz2005} (1D tubes) and in St\"oferle {\it et al.}~\cite{Stoferle2006} (3D lattices).

Lattice modulation spectroscopy, applicable to both bosons and fermion in lattices, was invented by Esslinger and coworkers in St\"oferle {\it et al.}~\cite{Stoferle2004} where it was used for studying the superfluid-Mott insulator transition in a 1D boson gas. In Ref.~\cite{Kollath2006} Kollath, Giamarchi and coworkers proposed that this method could be used
for fermions as well and theory description of the response in the fermion case was given. For an excellent detailed description of lattice modulation spectroscopy experiments see the book chapter by Tarruell \cite{Tarruell2015}. 

\begin{figure}
	\includegraphics[width=0.6\textwidth]{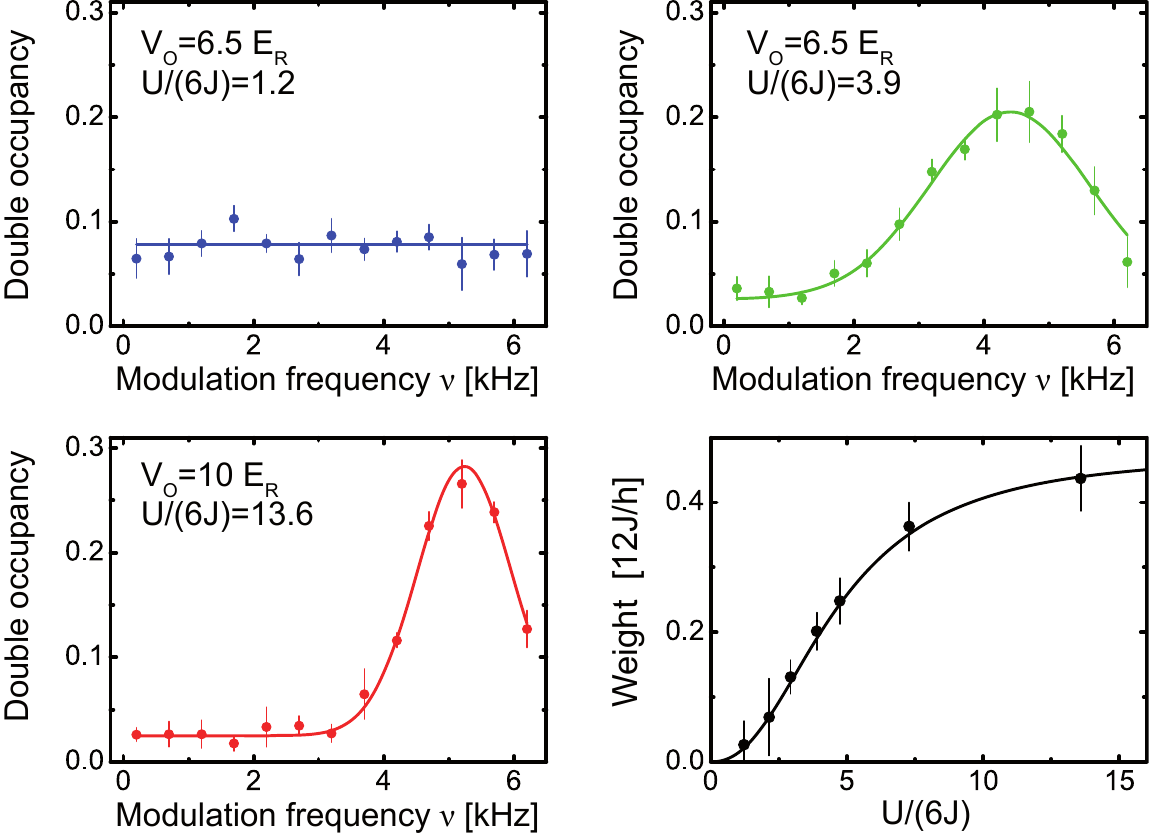}
	\caption{Lattice modulation spectroscopy was used in \cite{Jordens2008} to verify the existence of a gap, characteristic for the Mott state, for two-component repulsively interacting Fermions in optical lattices. In the Mott state, there is peak in the spectrum whose position and shape depend on the Mott gap given by $U$.} 
	\label{LatticeModulationImageTorma}
\end{figure}

Breakthroughs observing Mott insulators of fermions in lattices were achieved by J\"ordens, Esslinger and coworkers \cite{Jordens2008} and by Schneider, Bloch and coworkers \cite{Schneider2008}. Lattice modulation spectroscopy was used in \cite{Jordens2008} to identify the phases, see Figure \ref{LatticeModulationImageTorma}. Recently, even short range magnetic correlations have been observed by Greif, Esslinger and coworkers \cite{Greif2013}, and by Hart, Hulet and coworkers \cite{Hart2015}. In \cite{Hart2015}, spin-sensitive Bragg scattering of light was used to detect the correlations.   

\section{Universal relations and measuring the contact by the spectroscopies} \label{contactTorma}

In ultracold gases, the interparticle interactions have typically a short range but the  scattering length is large. Systems with such interactions show universal properties that depend 
only on the scattering length. The large scattering length may cause strong correlations, and it is often impossible to describe the system fully with theoretical 
methods, even non-perturbative ones. Nevertheless, {\it universal relations} that describe certain important aspects of the system can be derived. An important quantity in the universal relations 
is so-called {\it contact}. A lot of universal relations have been found for the two-component interacting Fermi gas. 
Many of the universal relations were first derived by Tan \cite{Tan2008,Tan20082,Tan20083}, therefore the name Tan's relations (Tan's contact) that is sometimes used for the universal relations (contact). Other types of early derivations are for instance by Braaten and Platter \cite{Braaten2008} and by Zhang and Leggett \cite{Zhang2009}. 
A useful review and pedagogical presentation of the topic is given in 
a book chapter by Braaten \cite{Braaten2012}. For a reader who wishes a briefer introduction to the topic I recommend Section 10.8. of my book chapter \cite{Torma2015}. 
 
The contact can be defined as \cite{Braaten2008}
\begin{equation}
C = \int d^3R \langle \Phi^\dagger \Phi (\mathbf{R}) \rangle , 
\end{equation}
where $\Phi (\mathbf{R}) = g_0 \psi_2 \psi_1  (\mathbf{R}) $ and $g_0$ is a cut-off dependent coupling constant. Clearly, the contact is related to the expectation value
of the interaction term in the Hamiltonian. It can be understood as a measure of the increased likelihood of finding two particles close (i.e., closer than
the scale of the scattering length) to each other due to 
the strong interactions. This means that in the momentum domain, it is a measure of atoms with large momentum. It is usually very difficult to evaluate the contact exactly. Fortunately, it is part of several universal relations which can each be experimentally verified and then compared. 

For instance, the adiabatic relation is a thermodynamic relation connecting the contact and the scattering length, expressed as
\begin{equation}
\left( \frac{dE}{da^{-1}}\right)_S = \left( \frac{dF}{da^{-1}}\right)_T = - \frac{\hbar}{4\pi m} C  .
\end{equation}
Here $E$ is the energy and $F=E - TS$ the free energy, $S$ the entropy and $a$ the scattering length.

Also spectroscopies can be used for measuring the contact.
The contact is related to the tail of the RF spectrum in the following way, in the case of a two-component gas with the RF transfer to a third state
that does not interact significantly with the two components (no final state interactions) \cite{Schneider2010}:
\begin{equation}
I(\omega) \rightarrow \frac{\Omega^2\sqrt{\hbar}}{4 \pi \sqrt{m} \omega^{3/2}} C , \label{RFtailContactTorma}
\end{equation}
where $C$ is the contact between the two species. The scaling of the decay with $\omega$ was also noted in Ref.~\cite{Pieri2009}, and can be obtained from BCS theory \cite{Torma2015}. 
For RF spectroscopy, a sum rule
involving the contact has also been derived \cite{Punk2007,Baym2007}
\begin{equation}
\int_{-\infty}^{\infty} \frac{d \omega'}{\pi} \omega I(\omega) = \frac{\hbar \Omega^2}{4m} \left( \frac{1}{a_{gg'}} 
- \frac{1}{a_{ge}}\right) C_{gg'}  .
\end{equation}

The structure factor as well has a tail that gives universal relations. Namely, for the dynamic structure factor
\begin{equation}
S(\omega, k) \rightarrow \frac{4q^4}{45\pi^2\omega (m\omega/\hbar)^{5/2}} \mathcal{C} ,
\end{equation}
and for the static one
\begin{equation}
S(k) \rightarrow \frac{1}{8}\left(\frac{1}{k}-\frac{4}{\pi a k^2}\right)\mathcal{C} . \label{staticTailTorma}
\end{equation}
Here $\mathcal{C}$ is the contact density (related to the contact by $C=\int d^3r \mathcal{C}(\mathbf{r})$).  

\begin{figure}
\includegraphics[width=0.7\textwidth]{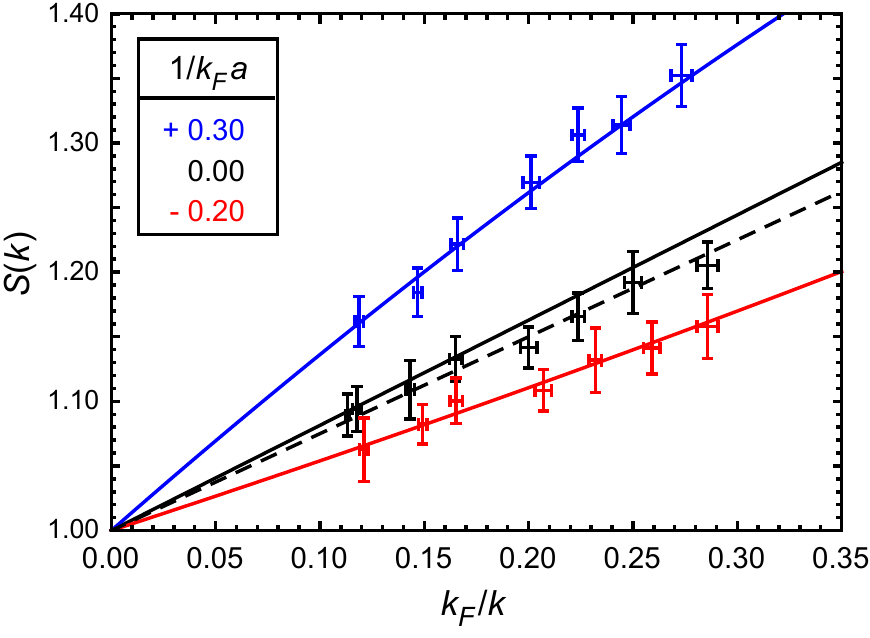}
\caption{The dependence of the static structure factor $S(k)$ on $k_F/k$ as measured in \cite{Hu2010}, for different interactions $1/(k_F a)$. The slope gives the contact according to Equation (\ref{staticTailTorma}) close to unitarity where the second term in the equation is small.}
	\label{ContactImageTorma}
\end{figure}

Universal relations have been verified in a large number of experiments. Here I mention only some related to spectroscopies. 
Kuhnle, Hoinka, Vale and coworkers used Bragg spectroscopy for verifying the universal relation for the tail of the static structure factor, 
Equation (\ref{staticTailTorma}) \cite{Hu2010,Hoinka2013}, see Figure \ref{ContactImageTorma}.
The contact was obtained by Stewart, Jin and coworkers from several different universal relations: from
measurement of the tail of the momentum distribution from ballistic expansion and by using the momentum-resolved RF spectroscopy (photoemission spectroscopy), and from
the tail of the RF spectrum, Equation (\ref{RFtailContactTorma}) \cite{Stewart2010}. These three independent 
measurements gave values of contact that were in good agreement with each other.   

As already discussed, the decaying tail of the RF spectrum is in striking contrast to the superconductor -- normal metal
experiments where the current continues to grow after the threshold. This is due to momentum conservation in RF spectroscopy. The momentum conservation leads to the fact that the RF spectrum
contains information about the spatial correlations, and since the pairing here was based on contact-interaction, 
the tail reflects the short-range nature of
the pairing. This makes intuitive the relation between the tail of the spectrum and the contact.  

\section{The question of a pseudogap versus a Fermi liquid}

One of the outstanding questions that ultracold Fermi gases may give the answer to is the following: what is the nature of the normal state of a strongly interacting Fermi system right above the superfluid critical temperature? Understanding the nature of the normal state is thought to provide a route for discovering even the underlying mechanism of high temperature superconductivity. There are differing theoretical predictions of what the state might be, ranging from a Fermi liquid to the so-called pseudogap state where a gap appears in the single particle excitation spectum already above $T_c$ but corresponds to incoherent Cooper pairs without long-range order. For more information, see for instance various chapters in the book edited by Zwerger \cite{Zwerger2012}, such as the one by Strinati \cite{Strinati2012}. For a description of how pseudogap theories may be probed both in ultracold gases and in cuprates, see the review by Chen, Levin and coworkers \cite{Chen2009}.  

The intriguing situation in the field of ultracold gases is now that there seems to be experimental evidence both for a pseudogap and a Fermi liquid state. Gaebler, Jin, Strinati, and coworkers presented an experiment \cite{Gaebler2010} that combined measurement of the condensate fraction to determine $T_c$ and momentum-resolved RF spectroscopy to measure the single particle spectral function. The observed back-bending of the spectral function around $k=k_F$ matched with a pseudogap theory by Perali, Pieri and Strinati (see also \cite{Perali2011}).   

Feld, K\"ohl and coworkers have reported evidence for pseudogap behaviour in {\it two-dimensional} Fermi gases \cite{Feld2011} using momentum-resolved RF spectroscopy (photoemission spectroscopy). Clear shifts of the RF peaks, caused by pairing effects, existed in the spectra well above any superfluid transition. However, the subtle difference between many-body pseudogap and two-body (no Fermi surface needed) pairing in 2D warrants further experimental and theoretical study \cite{Levinsen2015}. 

All the above reports on pseudogap behaviour were based on photoemission spectroscopy of $^{40}$K atoms. On the other hand, Fermi liquid description was found to be sufficient for desribing the strongly interacting normal state in the experiments of Nascimb{\`e}ne, Navon, Jiang, Chevy and Salomon \cite{Nascimbene2010,Navon2010} for $^6$Li atoms. They developed a method to extract the homogeneous equation of state from the density profiles of the trapped gas, which allowed obtaining thermodynamic quantities at high and low temperatures, for balanced and imbalanced gases. Also in the work of Sommer, Zwierlein and coworkers \cite{Sommer2011} the spin susceptibility of a gas of $^6$Li was found to agree with a Fermi liquid picture above $T_c$. 

Note that the choice of atom, $^{40}$K or $^6$Li, should not matter even when it seems to do so, because the behaviour of the gas is universal. The apparent differences may better be explained by noting that different atoms allow different probing techniques. For instance the momentum-resolved photoemission spectroscopy is well suited for $^{40}$K but not for $^6$Li. Different probing techniques can emphasize different features of the system. 

\begin{figure}
	\includegraphics[width=0.8\textwidth]{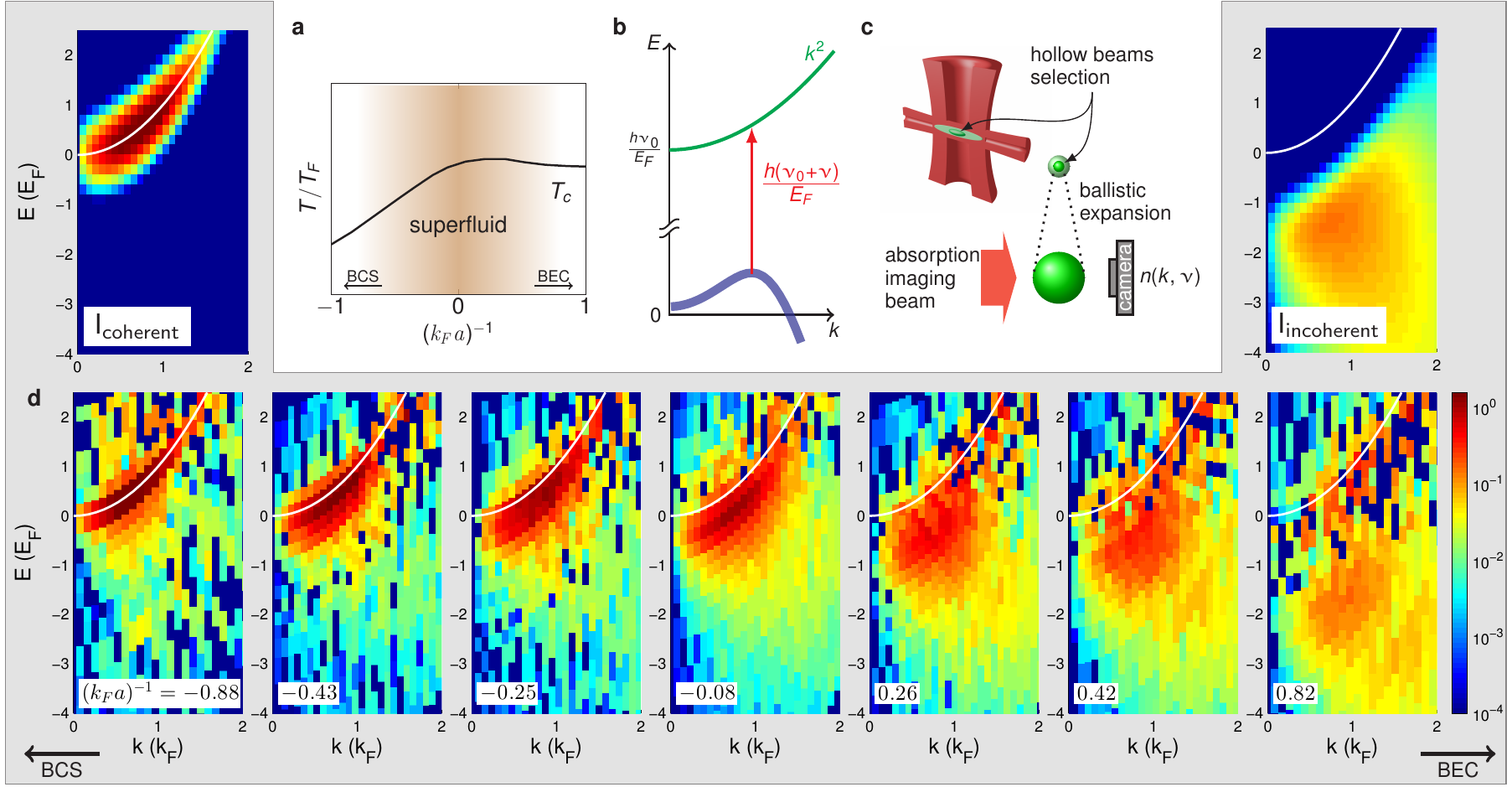}
	\caption{The evolution of momentum-resolved RF spectra (photoemission spectra) throughout the BCS-BEC crossover, from \cite{Sagi2015}. The evolution is well described by a combination of the coherent part (upper left panel) associated with a quasiparticle weight and an incoherent background (upper right panel). The schematic shows how only the centre of the trap was probed, in order to obtain results that correspond to a homogeneous system.} 
	\label{FermiLiquidImageTorma}
\end{figure}

Recently, Sagi, Jin and coworkers made an interesting experiment \cite{Sagi2015} with $^{40}$K where they removed trapping effects from the momentum-resolved RF spectroscopy response and then modelled it with a simple ansatz that contains a quasiparticle as given by Fermi liquid theory, together with an incoherent background that can describe the physics of pairing: 
\begin{equation}
I(k,\omega) = Z I_{coherent} (k, \omega) + (1-Z) I_{incoherent} (k, \omega) .
\end{equation}
The behaviour of the quasiparticle weight $Z$ is expected to depend on temperature \cite{Doggen2015}. The spectroscopy data was found to match well this description on the BCS side of the crossover and at unitarity, with the quasiparticle weight $Z$ disappearing on the BEC side at $1/(k_Fa) = 0.28(0.02)$. Thus Fermi liquid theory seems consistent with the experimental observations until a breakdown point, see Figure \ref{FermiLiquidImageTorma}. The article also presents comparisons of the results, for instance the Hartree energy extracted from the data, with quantum Monte-Carlo calculation by Magierski {\it et al.}~\cite{Magierski2009} and crossover theories by Haussmann, Punk and Zwerger \cite{Haussmann2009}, Kinnunen \cite{Kinnunen2012},  and  Bruun  and  Baym  \cite{Bruun2006}, with reasonably good match. The work of Haussmann {\it et al.}\ is based on Luttinger-Ward approach which does not include a pseudogap explicitly; neither does the Brueckner-Goldstone approach by Kinnunen where pairing is not considered but only scattering. On the other hand the work of Bruun and Baym was based on a Nozieres -- Schmitt-Rink approach where a pseudogap can be identified. 

Incorporating both the superfluid and the normal phase in a sufficiently correct manner to a crossover theory is a formidable theoretical challenge, and even the best crossover theories have their known pitfalls. The experiments have evolved enourmously, nevertheless they still have their limitations in precision and range of parameters that can be explored. Therefore I consider it still an extremely important future challenge for theory and experiment to nail down the nature of the normal state in a unitary Fermi gas --- although at the time of this writing, the Fermi liquid seems to have a slight lead.

\section{Outlook}

Further experiments and theory on the question whether the strongly interacting normal state is pseudogapped, Fermi liquid, or possibly something else is an obvious important direction of future research. Since the spectral function, gap, and single particle exctitations are essential for this question, spectroscopies are likely to turn out the be central in finding the answer.

One-dimensional systems \cite{Giamarchi2007} are likely to be an exciting arena to use spectroscopies, not only due to the physics to be expected but because the full field-matter dynamics of a many-body system can be exactly simulated \cite{Daley2004,Leskinen2010}. This allows to study also beyond linear response phenomena. 

Concerning many predicted superfluid and strongly correlated states in optical lattices, such as the antiferromagnetic state, the temperatures and entropies, have, however, not yet at the time of this writing reached low enough values. Spectroscopies, especially the lattice modulation and Bragg ones, may turn out to play a role in characterizing these states once low enough entropies are reached. The exotic 
Fulde--Ferrel--Larkin--Ovchinnikov (FFLO) state is predicted to be stabilized in lattices \cite{Koponen2007,Heikkinen2014}. In a one-dimensional lattice, it was predicted by exact simulations that the width of the lattice modulation spectra gives direct signatures of the FFLO state \cite{Korolyuk2010}. Also RF spectra may show in interesting ways the contributions of the non-paired quasiparticles in the FFLO state \cite{Bakhtiari2008}.

By recent breakthrough experiments \cite{Aidelsburger2013,Miyake2013,Jotzu2014,Aidelsburger2015}, the field of ultracold gases has entered the rich playground of topologically non-trivial quantum systems. Ultracold gases may be especially suited, for instance, to confirm the recently predicted connection between flat band high-$T_c$ superconductivity and the quantum geometric tensor and the Chern number \cite{Peotta2015}. How Chern numbers, edge currents, and the like might be characterized by various spectroscopies is a challenge both for theory and experiments. The future of spectroscopies in ultracold Fermi gases seems bright.    

\section*{Acknowledgements}
This  work  was  supported  by  the  Academy  of  Finland through its Centres of Excellence Programme (2012-2017)  and  under  Project  Nos.\ 263347,  251748,  and
272490,  and  by  the  European  Research  Council  
(ERC-2013-AdG-340748-CODE). Antti Paraoanu is thanked for producing Figure 1. The authors of the rest of the figures are thanked for providing the original images and for the kind permission to use them.  

\section*{References}
\bibliographystyle{iopart-num}
\bibliography{./YearOfLightTorma}
\end{document}